\documentclass[prb,twocolumn,showpacs,superscriptaddress]{revtex4}
\usepackage{array}
\usepackage{amsmath}
\usepackage{amssymb}
\usepackage{graphicx}
\usepackage[hypertex]{hyperref}
\usepackage{graphicx}
\usepackage{srctex}
\usepackage[usenames]{color}
\usepackage{dcolumn}
\usepackage{bm}

\definecolor{MidnightBlue}{cmyk}{0.98,0.13,0,0.43}
\definecolor{DarkGreen}{rgb}{0,0.7,0.1}

\begin{document}
\newcommand{\remark}[1] {\noindent\framebox{
\begin{minipage}{0.96\columnwidth}\textbf{\textit{ #1}}
\end{minipage}}
}

\newcommand{\bnabla}{{\boldsymbol{\nabla}}} \newcommand{\Tr}{\mathrm{Tr}} \newcommand{\Dk}{\check{\Delta}_{\cal K}} \newcommand{\Qk}{\check{Q}_{\cal K}} \newcommand{\Fk}{\check{\Phi}_{\cal K}}
\newcommand{\fk}{\check{\phi}_{\cal K}} \newcommand{\Ak}{\check{\mathbf{A}}_{\cal K}} \newcommand{\Si}{\check{\Xi}}
\newcommand{\cK}{\cal K}
\newcommand{\Lk}{\check{\Lambda}}
\newcommand{\bz}{{\mathbf z}}
\newcommand{\bx}{{\mathbf x}}
\newcommand{\br}{{\mathbf r}}
\newcommand{\bG}{{\mathbf G}}
\newcommand{\bu}{{\mathbf u}}
\newcommand{\bq}{{\mathbf q}}
\newcommand{\cH}{{\cal H}}
\newcommand{\dif}{{\mathrm d}}

\def\Xint#1{\mathchoice
{\XXint\displaystyle\textstyle{#1}}%
{\XXint\textstyle\scriptstyle{#1}}%
{\XXint\scriptstyle\scriptscriptstyle{#1}}%
{\XXint\scriptscriptstyle\scriptscriptstyle{#1}}%
\!\int}
\def\XXint#1#2#3{{\setbox0=\hbox{$#1{#2#3}{\int}$}
\vcenter{\hbox{$#2#3$}}\kern-.5\wd0}}
\def\ddashint{\Xint=}
\def\dashint{\Xint-}

\title{Duality between coherent quantum phase slip and Josephson junction in a nanosheet determined by the dual Hamiltonian method}
\author{M.~Yoneda}
\affiliation{Aichi University of Technology, 50-2 Umanori Nishihasama-cho, Aichi 443-0047, Japan}
\author{M.~Niwa}
\author{M.~Motohashi}
\affiliation{School of Engineering, Tokyo Denki University, 5 Senju Asahi-cho, Adachi-ku Tokyo\\ 120-855, Japan}

\date{\today}

\begin{abstract}
The duality between coherent quantum phase slip and Josephson junction in nanosheets was investigated using the dual Hamiltonian method. This is equivalent to the duality between superconductivity and superinsulator  in the 2 + 1 dimension at zero temperature. This method proved to be reliable within the Villain approximation. The possibility of the dual Ginzburg-Landau theory, which is the phenomenology of superinsulators and the dual BCS  theory, which is a microscopic theory, is also shown.
\end{abstract}



\maketitle
\def\thesection{\arabic{section}}
\setcounter{section}{0}
\section {Introduction}
In recent decades, two types of phenomena, which are considered to be dual states in superconductivity, have attracted attention. One of them is a phenomenon called coherent quantum phase slip, which is mainly known as a dual phenomenon of a Josephson junction ($J\!J$) in a one-dimensional superconducting system. The other is a phenomenon called superinsulator, which is mainly known as a dual phenomenon to superconductors (SC). A theoretical study of coherent quantum phase slip was submitted by Mooij et al.\cite{ref1}-\cite{ref3}  As flux quanta through nano superconducting wires, and an experimental demonstration of coherent quantum phase slip was performed by Astafiev et al.\cite{ref4} which embedded in a large superconducting loop with InOx Achieved in wire. The concept of superinsulator was first conceived in 1978 by 't Hooft\cite{ref5} as a theory explaining quark confinement. The superinsulator in the case of condensed matter was rediscovered by Diamantini et al.\cite{ref6}in 1996 as a periodic mixed Chern-Simons Abelian gauge theory describing charge-vortex  coupling equivalent to planar Josephson junction arrays in a self-dual approximation. Then, in 1998, Kr\"{a}mer and Doniach \cite{ref7} also rediscovered superinsulators using a phenomenological model in which vortices had a finite mass and traveled in a dissipative environment. Furthermore, in 2012, Yoneda et al. \cite{ref8} rediscovered superinsulators from a superinsulator / superconductor / superinsulator junction that is as the dual junction to Josephson junction. In previous studies, TiN\cite{ref9}, InO\cite{ref10} and NbTiN\cite{ref11} films have been experimentally observed to be superinsulating materials. Recently, superinsulators have attracted attention as a powerful desktop environment for QCD phenomena\cite{ref12,ref13} in order to realize a single-color version of quantum chromodynamics and elucidate quark confinement and asymptotic freedom. In a previous paper\cite{ref14}, we introduced two Hamiltonians dual to each other for a one-dimensional nanowire-based Josephson junction and a quantum phase slip junction ($Q\!P\!S\!J$), and applied dual conditions between the current and voltage of the electric circuit. A general theory to construct a dual system, called the dual Hamiltonian method, was proposed. Furthermore, a general discussion of the superconductor-superinsulator transition in a $1+1$ dimensional ($1+1d$) system at zero temperature from these two Hamiltonians was conducted. Our results show that in the $1+1d$ system at zero temperature, coherent quantum phase slip and superinsulator are completely equivalent phenomena. The main purpose of this work was to extend the theory of $1+1d$ systems on nanowires shown in the previous paper to the theory of $2+1d$ systems on nanosheets. The rest of this paper is organized as follows: In Section 2, we use the dual Hamiltonian method for between the JJ model and the $Q\!P\!S\!J$ model on a nanosheet at zero temperature to derive a between the phase and the amplitude dual relations, and a dual relations between the various constants.  In Section 3, based on the nonlinear Legendre transformation between the Lagrangians and the Hamiltonians with canonical conjugate variables of infinite order in a compact lattice space, and between the phase and the amplitude relationship derived in the previous section, we show that there is exact duality between the $J\!J$ model and the $Q\!P\!S\!J$ model. Additionally, the phase diagram between the $J\!J$ state (superconducting state) and the $Q\!P\!S$ state (superinsulating state) was discussed. In Section 4, we prove the validity of the results of the previous section by deriving a dual transformation from the anisotropic \scalebox{1.0}{$X\!Y(A\!X\!Y)$} model to the gauged dual anisotropic \scalebox{1.0}{$X\!Y(D\!A\!X\!Y)$} model by the Villain approximation in the $2+1d$ system. In Section 5, contrary to Section 4, we derive a dual transformation from the $D\!A\!X\!Y$ model to the $A\!X\!Y$ model with gauge by the Villain approximation in a $2+1d$ system. In Section 6, the dual Ginzburg-Landau ($D\!G\!L$) theory was derived from the mean field approximation of the gauged $Q\!P\!S\!J$ model, and the possibility of confinement of electric flux in the superinsulator was shown. In Section 7, we calculated the critical value of the $Q\!P\!S$ amplitude by the effective energy approach. In the summary and discussion in Section 8, we summarize the conclusions of this paper, discuss whether the minimum unit of charge confinement in a superinsulator is $2e$ or $e$, and, finally, we describe the possibility of dual BCS theory. In Appendix A, the numerical evaluation of the anisotropic massless lattice Green's function is shown at the origin of $x\!=\!0$, which is necessary for loop correction. In Appendix B, the effective energy approach for the $Q\!P\!S\!J$ model is explained.
\section{Dual Hamiltonian method between the Josephson junction (JJ) model and the quantum phase slip junction(QPSJ) model on a nanosheet at zero temperature \label{sec2}}
\begin{figure}[hbt]
\includegraphics[width=0.90\columnwidth]{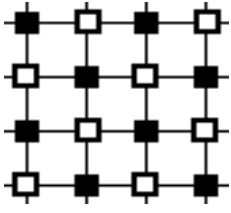}\\
\caption{A checkered nanosheet consisting of superconductors (black squares) and superinsulators (white squares).}
\end{figure}
Consider a checkered nanosheet consisting of superconductors and superinsulators as shown in Figure 1.
In such a $2d$ checkered nanosheet, when the superconductor region acts strongly, the $J\!J$ state becomes strong, and the Hamiltonian in this case is as follows:
\begin{align}\label{eq1}
\scalebox{0.85}{$\displaystyle
{H_{\!J\!J}}={E_c}\!\sum\limits_{\mathbf{x}=1}^{M}{\left[ N_{\theta }\left( x \right) \right]}^2+{E\!_J}\!\sum\limits_{\mathbf{x}=1}^M\sum\limits_{j=1}^2{\left[ 1-\cos {{\nabla\!}_j}\theta \left( x \right) \right]}
$},
\end{align}
where, \scalebox{1.0}{${N_{\theta }}\!\left(x\right)\!\equiv\! {N_{\theta }}\!\left( \mathbf{x},\tau  \right)$} means the number of the particles in the Cooper pair that is the canonical conjugate to the phase \scalebox{1.0}{$\theta$} of the Cooper pair, and the space difference\cite{ref14,ref15} of the phase \scalebox{0.9}{$\theta \left(x\right)\!\equiv\!\theta \left( \mathbf{x},\tau\right)$} is defined by \scalebox{0.9}{${{\nabla}_j}\theta\!\left(\mathbf{x},\tau \right)\!\equiv\!\theta\!\left( \mathbf{x},\tau\right)\!-\!\theta \left( \mathbf{x}\!-\!a\mathbf{e},\tau\right)$}; \scalebox{0.9}{$\tau\!\equiv\hbar \beta$} (\scalebox{0.9}{$\beta\!\equiv\!1/{{k_B}T}\;$}is the reverse temperature), $\mathbf{x}\!=\!\left(x_1, x_2\right)$ and $M\!\equiv\!{l^2}/{a^2}$ ($l$ and $a$are the space length and lattice spacing, respectively)are the imaginary time, the lattice coordinate points, and the total number of lattices in the two-dimensional lattice space, respectively; and ${E_c}\!\equiv\!{{\left(2e \right)}^2}/{2C}$ and ${E\!_J}\!\equiv\!{{\Phi_0}{I_c}}/{2\pi }$ are the charging energy per Cooper pair and the Josephson energy, respectively, where $C$,${I_c}$ and ${\Phi_0}\!=\!h/2e$ are the capacitance, the critical current and the magnetic flux-quantum, respectively. From the Hamiltonian of Eq. (\ref{eq2}), Josephson's equations are as follows:
\begin{align}\label{eq2}
\scalebox{0.9}{$\displaystyle
V\left( t \right)=i\frac{\hbar }{2e}\frac{\partial \theta \left( x \right)}{\partial \tau }=\frac{2{N_{\theta }}\left( x \right)}{2e}{E_c}
$},\nonumber \\
\scalebox{0.9}{$\displaystyle
{I_j}\left(x\right)=-\frac{2\pi }{{\Phi_0}}{E_J}\sin {\nabla_j}\theta \left( x \right).\;
$}
\end{align}
where, $V\left( x \right)$ and ${I_j}\left( x \right)$ ($j \!=\!1, 2$) are the voltage and the $j$ components of the current in the $J\!J$, respectively. On the other hand, when the superinsulator region acts strongly, the $Q\!P\!S\!J$ state becomes strong, and the Hamiltonian in this case is as follows:
\begin{align}\label{eq3}
\scalebox{0.85}{$\displaystyle
{H_{Q\!P\!S}}={E_L}\sum\limits_{\mathbf{x}=1}^M{{{\left[ {\tilde{N}_{\tilde{\theta }}}\left( x \right) \right]}^2}}+{E_S}\sum\limits_{j=1}^2{\sum\limits_{\mathbf{x}=1}^M{\left[ 1-\cos {{\nabla }_j}\tilde{\theta }\left( x \right) \right]}}
$},
\end{align}
where, ${{\tilde{N}}_{\tilde{\theta }}}\left( x \right)\equiv{\tilde{N}_{\tilde{\theta }}}\left( \mathbf{x},\tau  \right)$ means the magnetic flux quantum numbers $\tilde{\theta }\left( x \right)\equiv \tilde{\theta }\left( \mathbf{x},\tau  \right)$ of the magnetic flux quantum field. $\,{E_L}\equiv {{{\Phi }_{0}}^2}/{2L}\;$ and $\,{E_S}\equiv {2e{V_c}}/{2\pi }\;$ are the inductive energy per magnetic flux quantum and the $Q\!P\!S$ amplitude, respectively, where $\,{V_c}$ is the critical voltage. From the Hamiltonian of Eq.(\ref{eq3}), the dual Josephson equations are as follows:
\begin{align}\label{eq4}
\scalebox{0.9}{$\displaystyle
\tilde{V}\left( x \right)=i\frac{\hbar }{{\Phi }_0}\frac{\partial \tilde{\theta }\left( x \right)}{\partial \tau }=\frac{2{\tilde{N}_{\tilde{\theta }}}\left( x \right)}{\Phi _0}{E_L}
$},\nonumber \\
\scalebox{0.9}{$\displaystyle
{\tilde{I}_j}\left( x \right)=\frac{2\pi }{2e}{E_S}\sin {{\nabla }_j}\tilde{\theta }\left( x \right),\;\;\;\;\;\;\;\;
$}
\end{align}
where, $\tilde{V}\left( x \right)$ and ${\tilde{I}_j}\left( x \right)$ ($j \!=\!1, 2$) are the dual voltage and the j components of dual current in the $Q\!P\!S\!J$, respectively. As the first step of the dual Hamiltonian method, two dual conditions\cite{ref8,ref14} between Eq.(\ref{eq2}) and the dual equations of Eq.(\ref{eq4}) are assumed as follows:
\begin{align}\label{eq5}
\scalebox{0.9}{$\displaystyle
V\left( x \right)\equiv \tilde{I}\left( x \right),\text{ }I\left( x \right)\equiv \tilde{V}\left( x \right),
$}.
\end{align}
where, $I\!\equiv\!\sqrt{{I_1}\!^2+{I_2}\!^2}$ and $\tilde{I}\!\equiv\!\sqrt{{\tilde{I}_1}^2+{\tilde{I}_2}^2}$ are the intensity of the current in the $J\!J$ and the intensity of the dual current in the $Q\!P\!S\!J$, respectively. When the condition of Eq.(\ref{eq5}) is imposed between Eqs.(\ref{eq2}) and (\ref{eq4}), the following two relational expressions between the phase and the number of particles between dual systems are derived as shown below. One of them is the relationship between the phase $\theta$ of the Cooper pair and the magnetic flux quantum numbers ${\tilde{N}_{\tilde{\theta }}}$, and the other is the relationship between the phase $\tilde{\theta }$ of the magnetic flux quantum and the number ${N_{\theta }}$ of the Cooper pair, as follows:
\begin{align}\label{eq6}
\scalebox{0.85}{$\displaystyle
{{N}_{\theta }}\left( x \right)=\frac{1}{2\pi }\sqrt{\sum\limits_{j=1}^2{{{\sin }^2}{\nabla _j}\tilde{\theta }\left( x \right)}}
$},\nonumber \\
\scalebox{0.85}{$\displaystyle
{{\tilde{N}}_{\tilde{\theta }}}\left( x \right)=\frac{1}{2\pi }\sqrt{\sum\limits_{j=1}^2{{{\sin }^2}{\nabla _j}\theta \left( x \right)}}
$},
\end{align}
In Eq.(\ref{eq6}), the leftmost equation shows that the Cooper pair number is proportional to the strength of the flux quantum current, and the next equation Is shown that the flux quantum number is proportional to the strength of the Cooper pair current. Since the Cooper pair current and the flux quantum current form a closed loop, the number of Cooper pairs and the flux quantum number can be considered as the winding number of the current loop formed from each dual current. If the relationships described in Eq.(\ref{eq6})  are satisfied, the relationship between the $Q\!P\!S$ amplitude and the charging energy per single charge, and the relationship between the Josephson energy and the inductive energy per magnetic flux-quantum, are as follows:
\begin{align}\label{eq7}
\scalebox{0.95}{$\displaystyle
{E_S}=\frac{E_c}{2{\pi }^2},\;\;\;                         
{E_J}=\frac{E_L}{2{\pi }^2}
$}. 
\end{align} 
Furthermore, the inductance and capacitance are related to the critical current and the critical voltage, respectively, as follows:
\begin{align}\label{eq8}
\scalebox{0.95}{$\displaystyle
L=\frac{\Phi _0}{2\pi {I_c}},\;\;\;         
C=\frac{2e}{2\pi {V_c}}
$},
\end{align} 
The Lagrangians at zero temperature in Eqs.(\ref{eq1}) and Eq.(\ref{eq2}) are as follows:
\begin{align}\label{eq9}
\scalebox{0.8}{$\displaystyle
{L\!_{J\!J}}\!=\!-\!\sum\limits_{\mathbf{x}=1}^{M}\!{\left\{ \frac{E\!_J^0}{2}{{\left[ {{\nabla }_{\tau }}\theta \left( x \right) \right]}^2}\!+\!{E\!_J}\!\sum\limits_{j=1}^2{\left[ 1-\cos {{\nabla }_j}\theta \left( x \right) \right]} \right\}},\;\;\;
$}
\end{align} 
\begin{align}\label{eq10}
\scalebox{0.8}{$\displaystyle
{L\!_{Q\!P\!S}}\!=\!-\!\sum\limits_{\mathbf{x}=1}^{M}\!{\left\{ \frac{E\!_S^0}{2}{{\left[ {{\nabla }_{\tau }}\tilde{\theta }\left( x \right) \right]}^2}\!\!+\!{E\!_S}\!\sum\limits_{j=1}^2{\left[ 1-\cos {{\nabla }_j}\tilde{\theta }\left( x \right) \right]} \right\}}
$},
\end{align} 
where $E_J^0$ and $E_S^0$ can be considered as the imaginary time components of the $J\!J$ and the $Q\!P\!S\!J$, respectively, and are defined as follows:
\begin{align}\label{eq11}
\scalebox{0.9}{$\displaystyle
E_J^0\equiv \frac{{\hbar }^2}{2{a_0}^2{E_c}}, \;\;\; E_{S}^0\equiv \frac{{\hbar }^2}{2{a_0}^{2}{E_L}},
$}
\end{align} 
where ${a_0}\equiv {\tau }_{\max }/M_{\tau }$ is an imaginary time spacing in the time dimension; and ${\tau }_{\max }$ and $M_{\tau }$  are an imaginary time length and an imaginary time division number, respectively. Eqs.(\ref{eq7}) and (\ref{eq11}) can be summarized as a relationship between the Josephson energy and the $Q\!P\!S\!J$ energy as follows:
\begin{align}\label{eq12}
\scalebox{0.9}{$\displaystyle
{E'}_J^0\equiv \frac{1}{4{{\pi }^2}{{E'}\!_S}}, \;\;\; {E'}_S^0\equiv \frac{1}{4{{\pi }^2}{{E'}\!_J}},
$}
\end{align} 
where ${{E'}\!_J}\equiv {{E_J}{a_0}}/{\hbar }$, ${E'}\!_J^0\equiv {E_J^0{a_0}}/{\hbar }$, ${{E'}\!_S}\equiv {{E_S}{a_0}}/{\hbar }$ and ${E'}\!_S^0\equiv {E_S^0{a_0}}/{\hbar }$ represent the nondimension energies, respectively.The first terms of Eq.(\ref{eq9}) and (\ref{eq10})  are expressed in a quadratic form for the imaginary time difference of each phase, however, the rightmost terms in Eqs.(\ref{eq9}) and (\ref{eq10}) are expressed in a cosine form for the spatial difference of each phase. Here, the cosine form also introduces approximately to the leftmost terms of Eqs.(\ref{eq9}) and (\ref{eq10}) in consideration of the periodicity in the lattice space as follows:
\begin{align}\label{eq13}
\scalebox{0.75}{$\displaystyle
{L\!_{A\!X\!Y}}\!=\!-\!\sum\limits_{\mathbf{x}=1}^M{\left\{\!E_J^0\Bigl[1-\cos {{\nabla }_{\tau }}\theta \left( x \right) \Bigr]\!+\!{E_J}\sum\limits_{j=1}^2{\Bigl[1-\cos {{\nabla }_j}\theta \left( x \right) \Bigr]} \right\}}
$},
\end{align} 
\begin{align}\label{eq14}
\scalebox{0.75}{$\displaystyle
{L_{D\!A\!X\!Y}}\!=\!-\!\sum\limits_{\mathbf{x}=1}^M{\left\{\!E_S^0\left[1-\cos {{\nabla }_{\tau }}\tilde{\theta }\left( x \right) \right]\!+\!{E_S}\sum\limits_{j=1}^2{\left[1-\cos {{\nabla }_x}\tilde{\theta }\left( x \right) \right]} \right\}}
$}.                   
\end{align}
where, ${L_{A\!X\!Y}}$ and ${L_{D\!A\!X\!Y}}$ are equivalent to ${L_{J\!J}}$ and ${L_{Q\!P\!S}}$, respectively, and the $A\!X\!Y$ and $D\!A\!X\!Y$ model, respectively. From the Lagrangian in Eq.(\ref{eq13}) the partition function of the $J\!J$ state (superconducting state) at zero temperature is as follows:
\begin{align}\label{eq15}
\scalebox{0.8}{$\displaystyle
Z\!_{A\!X\!Y}\equiv\exp\left\{ -\Bigl({E^{\prime }}\!_J^0+{E^{\prime }}\!_J\Bigr)\!M{M_{\tau }} \right\}{{Z'}\!_{A\!X\!Y}}
$},\quad\quad\;\;\;\nonumber\\ 
\scalebox{0.8}{$\displaystyle
{Z'}\!_{A\!X\!Y}\!\!\equiv\!\!\int\!\!{D\theta }\exp\!\sum\limits_{x,\tau }\!{\left[ {E^{\prime }}_J^0\cos {\nabla\!_{\tau }}\theta \left( x \right)+{E^{\prime }}\!_J\sum\limits_{j=1}^2{\cos {\nabla\!_j}\theta\left( x \right)} \right]}
$}.                      
\end{align}
where $\sum\limits_{x}\!\!\!\equiv\!\!\sum\limits_{\tau\!=\!1}^{M_{\tau }}\!{\sum\limits_{\mathbf{x}\!=\!1}^M}$ and $\int\!\!{D\theta\!}\!\!\equiv\!\!\prod\limits_{\tau\!=\!1}^{M_{\tau }}{\!\prod\limits_{\mathbf{x}\!=\!1}^{M}{\!\!\int\limits_{-\pi }^{\pi }\!\!{\frac{d\theta\left(\mathbf{x},\tau  \right)}{2\pi }}}}$ are sums and path integrals, respectively, in $2+1d$ lattice space $x\!=\!(x, \tau)$. Similarly, from the Lagrangian in Eq.(\ref{eq14}), the partition function of the $Q\!P\!S\!J$ state (superinsulating state) at zero temperature is as follows:
\begin{align}\label{eq16}
\scalebox{0.8}{$\displaystyle
{Z_{D\!A\!X\!Y}}\equiv\exp\left\{ -\!\left( {E^{\prime }}_S^0+{E^{\prime }}_S\right)\!{M_{\tau }}{M_x} \right\}{{Z'}\!_{D\!A\!X\!Y}}
$},\quad\quad\nonumber\\ 
\scalebox{0.8}{$\displaystyle
{{Z'}\!_{D\!A\!X\!Y}}\!\!\equiv\!\!\int\!\!{D\tilde{\theta }}\exp\!\sum\limits_x\!{\left[ {E^{\prime }}_S^0\cos {\nabla\!_{\tau }}\tilde{\theta }\left( x \right)+{E^{\prime }}\!_S\!\sum\limits_{j=1}^2{\cos {\nabla\!_j}\tilde{\theta }\left( x \right)} \right]}
$}.      
\end{align}
The partition functions $Z_{A\!X\!Y}$ and $Z_{D\!A\!X\!Y}$  in Eqs.(\ref{eq15}) and (\ref{eq16}) are the starting points for the analysis in the following sections.
\section{Dual transformation between the JJ model and the QPSJ model on the nanosheet \label{sec3}}
This section describes the nonlinear Legendre transformation between the Hamiltonian and the Lagrangian with canonical conjugate variables of the infinite order in a compact $2+1d$ lattice space on a nanosheet. Herein, we show that there is exact duality between the $J\!J$ model and the $Q\!P\!S\!J$ model. For the partition functions ${Z\!_{J\!J}}$ in the Lagrangian of Eq.(\ref{eq9}), the auxiliary field $N\left( x \right)$ by Hubbard-Stratonovich transformation is introduced as follows: 
\begin{align}\label{eq17}
\scalebox{0.9}{$\displaystyle
Z\!_{J\!J}\!=\!\!\!\int\!\!\!{D\theta }\!\!\!\int\!\!\!{D{N}\!}\exp\!\sum\limits_x\!{\Bigl\{ iN\!\left( x \right)\!{\nabla\!_{\tau }}\theta \left( x\right)\!-\!{{E'}\!_C}N{{\left( x \right)}^2}
\Bigr.}$}\;\;\nonumber\\
\scalebox{0.9}{$\displaystyle
\Bigl.  
-\!{{E'}\!_J}\!\sum\limits_{j=1}^2{\left[ 1\!-\!\cos {{\nabla}\!_j}\theta \left( x \right) \right]}
\Bigr\}
 $},\quad
\end{align}
It can be seen that the auxiliary field $N\left( x \right)$ is the same as the number ${N_{\theta }}\left( x \right)$ of the Cooper pairs introduced in Eq.(\ref{eq1}).The canonical conjugate momentum ${{p}_{\theta }}\left( x \right)$ with respect to $\theta \left( x \right)$ in the Lagrangian of the lattice space of Eq.(\ref{eq9}) is defined as follows:.
\begin{align}\label{eq18}
\scalebox{0.98}{$\displaystyle
i{p_\theta}\!\left( x \right)\!\equiv\!i\hbar {N}\!\left( x \right)\!=\!-{a_0}E_J^0{\nabla_{\tau }}\theta \left( x \right)
$}.  
\end{align}
On the other hand, for the partition functions $Z_{A\!X\!Y}$ in Eq.(\ref{eq15}), the auxiliary field $N\!\left( x \right)$ by Hubbard–Stratonovich transformation is introduced as follows:
\begin{align}\label{eq19}
\scalebox{0.83}{$\displaystyle
Z\!_{A\!X\!Y}\!=\!\!\int\!\!{D\theta }\!\!\!\int\!\!{DN\!}\exp\!\sum\limits_x\!{\Bigl\{ i2N\left( x \right)\sin \left[ {\nabla\!_{\tau }\theta \left( x \right)}/2\right]\!-\!{{E'}\!_C}N{{\left( x \right)}^2}
\Bigr.}$}\nonumber\\
\scalebox{0.83}{$\displaystyle
\Bigl.  
-{{E'}\!_J}\sum\limits_{j=1}^2{\left[ 1\!-\!\cos {\nabla\!_j}\theta\!\left( x \right) \right]}
\Bigr\}
 $},\quad
\end{align}
Again, it can be seen that the auxiliary field $N\left( x \right)$ is the same as the number ${N_{\theta }}\left( x \right)$ of the Cooper pairs introduced in Eq.(\ref{eq1}). The canonical conjugate momentum ${{p}_{\theta }}\!\left( x \right)$ with respect to $\theta\!\left( x \right)$ in the Lagrangian of the compact lattice space of Eq.(\ref{eq13}) is defined as follows:
\begin{align}\label{eq20}
\scalebox{0.95}{$\displaystyle
i{p_{\theta }}\left( x \right)\equiv i\hbar {N_{\theta }}\left( x \right)=-2{a_0}E_J^0\sin \Bigl[ {\nabla _{\tau }\theta \left( x \right)}/2 \Bigr]
$},
\end{align}                  
In Eq.(\ref{eq20}), if the linear approximation of $\sin\!\left({\nabla\!_{\tau }\theta\!}/2\right)\!\!\simeq\!\!{{\nabla _{\tau }}\!\theta\!}/2$ holds, it matches the canonical conjugate momentum in Eq.(\ref{eq18}).Therefore, the contents of the curly bracket of the exponential function between Eqs.(\ref{eq15}) and (\ref{eq19}) can be considered as a “nonlinear Legendre transformation” introduced by the canonical conjugate momentum of Eq.(\ref{eq20}). Similarly, for the partition function $Z\!_{D\!A\!X\!Y}$ in Eq.(\ref{eq16}), the auxiliary field $\tilde{N}\left( x \right)$ by Hubbard–Stratonovich transformation is introduced as follows:
\begin{align}\label{eq21}
\scalebox{0.83}{$\displaystyle
Z\!_{D\!A\!X\!Y}\!=\!\!\int\!\!\!{D\tilde{\theta }}\!\!\!\int\!\!\!{D\tilde{N}\!}\exp\!\sum\limits_x\!{\Bigl\{ i2\tilde{N}\!\left( x \right)\sin\!\left[ {\nabla\!_{\tau }\tilde{\theta }\!\left( x \right)}/2\right]\!\!-\!{{E'}\!_L}\tilde{N}{{\left( x \right)}^2}
\Bigr.}$}\nonumber\\
\scalebox{0.83}{$\displaystyle
\Bigl.  
-{{E'}\!_S}\sum\limits_{j=1}^2\!{\left[ 1\!-\!\cos {\nabla\!_j}\tilde{\theta}\!\left( x \right) \right]}
\Bigr\}
 $},\quad
\end{align}
It can be seen that the auxiliary field $\tilde{N}\left( x \right)$ is the same as the magnetic flux quantum numbers $\tilde{N}_{\tilde{\theta }}\left( x \right)$ in Eq.(\ref{eq3}). The canonical conjugate momentum ${\tilde{p}_{\tilde{\theta }}}\left( x \right)$ with respect to $\tilde{\theta }\left( x \right)$ in the Lagrangian of the compact lattice space of Eq.(\ref{eq14}) is defined as follows:
\begin{align}\label{eq22}
\scalebox{0.95}{$\displaystyle
i{\tilde{p}_{\tilde{\theta} }}\left( x \right)\equiv i\hbar {\tilde{N}_{\tilde{\theta} }}\left( x \right)=-2{a_0}E_S^0\sin\!\left[ {\nabla _{\tau }\tilde{\theta}\left( x \right)}/2 \right]
$},
\end{align}          
Therefore, the contents of the curly brackets of the exponential function between Eqs.(\ref{eq16}) and (\ref{eq21}) can also be considered as the“nonlinear Legendre transformation” introduced by the canonical conjugate momentum of Eq.(\ref{eq22}). Since Eqs.(\ref{eq20}) and (\ref{eq22}) have a half-angle relationship, a half-angle version of Eq.(\ref{eq6}) is introduced as follows:
\begin{align}\label{eq23}
\scalebox{0.82}{$\displaystyle
{N_{\theta }}\left( x \right)=\frac{1}{2\pi }\sqrt{\sum\limits_{j=1}^2{{\sin }^2\left[ {{\nabla _j}\tilde{\theta }\left( x \right)}/2 \right]}}
$},\nonumber \\
\scalebox{0.82}{$\displaystyle
\tilde{N}_{\tilde{\theta }}\left( x \right)=\frac{1}{2\pi }\sqrt{\sum\limits_{j=1}^2{{\sin }^2\Bigl[ {{\nabla _j}\theta \left( x \right)}/2 \Bigr]}}
$},
\end{align}
Eq.(\ref{eq23}) is equivalent to Eq.(\ref{eq6}) within the range of linear approximation.  From Eqs.(\ref{eq20}), (\ref{eq21}), and (\ref{eq23}), the relationship between the phase $\theta \left( x \right)$ of the Cooper pair and the phase $\tilde{\theta }\left( x \right)$ of the magnetic flux quantum field is as follows:
\begin{align}\label{eq24}
\scalebox{0.82}{$\displaystyle
N_{\theta }^2\!\left( x \right)\!=\!-\frac{4{\left( {a_0}E_J^0 \right)}^2}{\hbar^2}{\sin^2}\Bigl[ \nabla_{\tau }\theta \left( x \right)/2 \Bigr]\!=\!\frac{1}{\pi^2}\!\sum\limits_{j=1}^2{{\sin^2}\left[ {{\nabla }_j\tilde{\theta }\left( x \right)}/2 \right]}
$},\nonumber \\
\scalebox{0.82}{$\displaystyle
\tilde{N}_{\tilde{\theta }}^2\!\left( x \right)\!=\!-\frac{4{\left( {a_0}E_S^0 \right)}^2}{\hbar^2}{\sin^2}\left[ {\nabla _{\tau }\tilde{\theta }\left( x \right)}/2 \right]\!=\!\frac{1}{\pi^2}\!\sum\limits_{j=1}^2{{\sin ^2}\Bigl[ {\nabla_j}\theta \left( x \right)/2 \Bigr]}
$},
\end{align} 
Using Eqs.(\ref{eq7}), (\ref{eq11}) and (\ref{eq24}), the following relationships are  obtained:
\begin{align}\label{eq25}
\scalebox{0.75}{$\displaystyle
E_c\!\sum\limits_{\mathbf{x}}{{\left[ N_{\theta }\left( x \right) \right]}^2}\!\!=\!\!-E_J^0\sum\limits_{\mathbf{x}}\!{\Bigl[1\!-\!\cos {\nabla _{\tau }}\theta \left( x \right) \Bigr]}\!=\!E_S\!\sum\limits_{\mathbf{x}}{\sum\limits_{j=1}^2{\left[ 1\!-\!\cos {\nabla _j}\tilde{\theta }\left( x \right) \right]}}
$},
\end{align}    
\begin{align}\label{eq26}
\scalebox{0.75}{$\displaystyle
E_L\!\sum\limits_{\mathbf{x}}{{\left[ \tilde{N}_{\tilde{\theta }}\left( x \right) \right]}^2}\!\!\!\!=\!\!-E_S^0\sum\limits_{\mathbf{x}}\!{\left[ 1\!-\!\cos {\nabla _{\tau }}\tilde{\theta }\left( x \right) \right]}\!=\!E_J\sum\limits_{\mathbf{x}}{\sum\limits_{j=1}^2{\Bigl[1\!-\!\cos {{\nabla }_j}\theta \left( x \right) \Bigr]}}
$}.
\end{align}    
Eq.(\ref{eq25}) means that the charging energy by the charge $2e$ in the $J\!J$ is equal to the $Q\!P\!S\!J$ energy, that is, the condensation energy of the magnetic flux. Eq.(\ref{eq26}) means that the charging energy by the flux quantum $\Phi_0$ in the $Q\!P\!S\!J$ is equal to the $J\!J$ energy, that is, the condensation energy of the Cooper pair. By establishing the relationship between Eqs.(\ref{eq25}) and (\ref{eq26}), between the Hamiltonian Eqs.(\ref{eq1}) and (\ref{eq3}), and, between the Lagrangian Eqs.(\ref{eq13}) and (\ref{eq14}), it can be shown that each of them is self-dual. Thus, from the canonical conjugate momentum of Eqs.(\ref{eq20}) and (\ref{eq22}), and the relationship between the phase and amplitude of Eq.(\ref{eq6}) derived in the previous section, an exact duality between the $J\!J$ and $Q\!P\!S\!J$ models has been proven. From the Josephson's equations Eq.(\ref{eq2}) and the dual Josephson's equations Eq.(\ref{eq4}), the electrical conductance $G$ can be derived as follows\cite{ref8,ref14}:
\begin{align}\label{eq27}
\scalebox{0.85}{$\displaystyle
G=-{G_Q}\frac{d{N_{\theta }}\left( x \right)}{d{\tilde{N}_{\tilde{\theta }}}\left( x \right)}={G_Q}\frac{E_J}{E_S}\frac{{\tilde{N}_{\tilde{\theta }}}\left( x \right)}{N_{\theta }\left( x \right)}
  $},
\end{align} 
 where ${G_Q}\equiv {{\left( 2e \right)}^2}/{h}$ is the quantum conductance. From Eq.(\ref{eq27}), the following elliptical orbit can be drawn:
\begin{align}\label{eq28}
\scalebox{1.0}{$\displaystyle
 g^{-1}{N_{\theta }}^2+g\tilde{N}_{\tilde{\theta }}^2=\frac{1}{\eta }
$},\;\;\;\;\;\;\;\;\;\;\;\nonumber \\
\scalebox{1.0}{$\displaystyle
g\equiv {\left( E\!_J/E\!_S \right)}^{1/2},\;\eta \equiv \sqrt{{E\!_J}{E\!_S}}/{\Gamma }
$},
\end{align} 
where $\Gamma$ is an arbitrary integral constant  with an energy dimension. From Eq.(\ref{eq27}), one can draw an elliptical orbit as shown in Figure 2.
\begin{figure}[htbp]
 \begin{minipage}{1.0\hsize}
  \begin{center}
   \includegraphics[keepaspectratio, height=38mm]{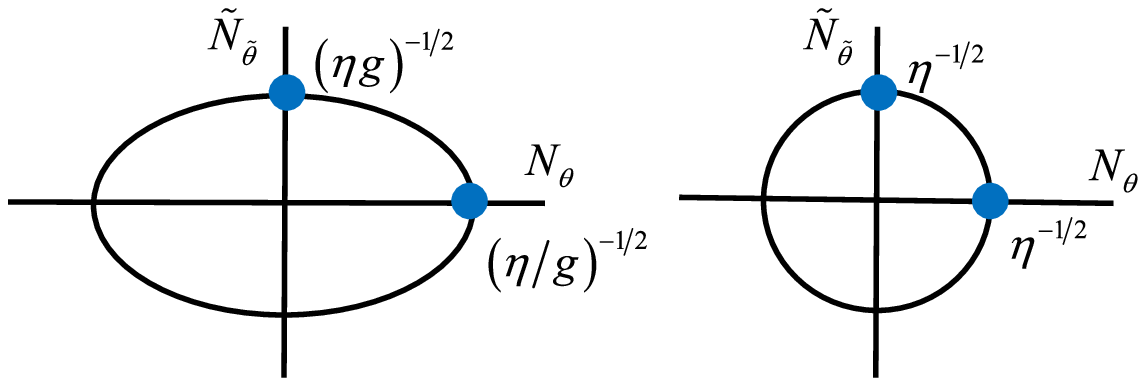}
   \begin{tabular}{cccc}
\;\;\;\;\;\;\;\;\;(a) $E_J\gg E_S$&\qquad\qquad\qquad\;\;(b) $E_J=E_S$\\
\end{tabular}
  \end{center}
\label{fig:one}
 \end{minipage}\\
 \begin{minipage}{1.0\hsize}
\begin{center}
 \includegraphics[keepaspectratio, height=38mm]{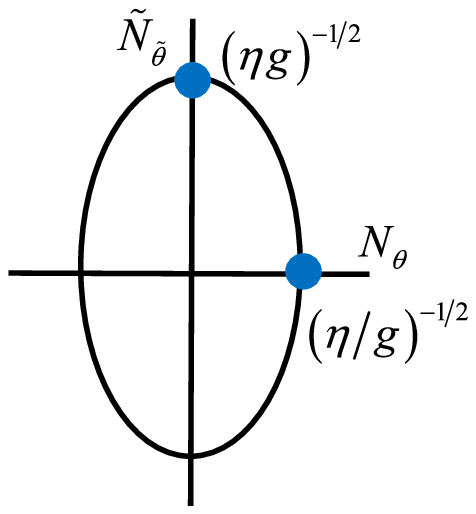}
  \begin{tabular}{cccc}
\;(c) $E_J\ll E_S$
\end{tabular}
 \end{center}
\caption{Elliptical orbit with the number ${N_{\theta }}$ of the Cooper pairs versus the number ${\tilde{N}_{\tilde{\theta }}}$ of the magnetic flux quantum }
\label{fig:two}
 \end{minipage}
\end{figure}
Figure 2 (a) shows a $J\!J$ junction state (superconducting state) for $E_J\gg E_S$, Figure 2 (b) shows a quantum Hall state (Bose semiconductor state) for $E_J=E_S$, and Figure 2 (c) shows a $Q\!P\!S$ junction state (superinsulating state) for $E_J\ll E_S$.The results shown in Figure 2 are very similar to those in references\cite{ref12}'\cite{ref15}'\cite{ref16}. Compare the result of Eq.(\ref{eq27}) with the result of the quantum Hall effect shown below:.   
\begin{align}\label{eq29} 
\scalebox{0.85}{$\displaystyle
G={G_Q}\nu 
$},
\end{align}
where $\nu$ is the Landau level occupancy and is defined as:
\begin{align}\label{eq30} 
\scalebox{0.85}{$\displaystyle
\nu \equiv \frac{N_e}{N_{\phi }}
$},
\end{align} 
where $N_e$ and $N_{\theta}$ are the electron number and the number of magnetic fluxes (vortex number), respectively. In our model, the correspondence is ${N_e}\!\to\!N_{\theta }$ and $N_{\phi }\!\to\!\tilde{N}_{\tilde{\theta }}$. When the Hall conductivity of the quantum Hall effect in the bulk is derived using the Kubo formula, the Landau level occupancy $\nu$ can also be expressed as the topological quantum number, also called the Chern number. From Eqs.(\ref{eq23}), (\ref{eq27}), (\ref{eq29}) and (\ref{eq30}), it can be seen that the quantum Hall state is obtained if the following relational expression is satisfied:
\begin{align}\label{eq31}
\scalebox{0.85}{$\displaystyle
E_J\Bigl[ 1-\cos \theta \left( x \right) \Bigr]=E_S\left[ 1-\cos \tilde{\theta }\left( x \right) \right]
$},
\end{align}
Eq.(\ref{eq31}) means that the energy of the Josephson junction is equal to the energy of the quantum phase slip. In the vicinity of the region where the Landau level occupancy $\nu$ is represented by the Chern number, it is expected that not only the quantum Hall phase but also a topological insulator phase and a topological metal phase exist. If $\eta>1$ and $g\!\equiv\!\left(E_J/E_S\right)^{1/2}\!\leqq\!1$ , the region means Bose insulator\cite{ref16}-\cite{ref19}, and if $\eta>1$ and $g\equiv\left(E_J/E_S\right)^{1/2}\geqq1$, the region means Bose metal\cite{ref16,ref20}-\cite{ref24}. Figure 3 shows a schematic phase diagram for $g$ versus $\eta$.
\begin{figure}[htbp]
 \begin{minipage}{1.0\hsize}
 \includegraphics[keepaspectratio, height=35mm]{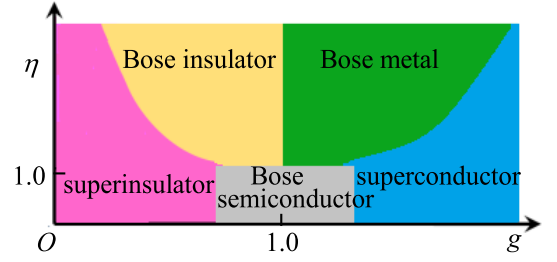}
\caption{Schematic phase diagram for $g\equiv {\left( E\!_J/E\!_S \right)}^{1/2}$-$\eta \equiv \sqrt{{E\!_J}{E\!_S}}/{\Gamma }$. Schematic phase diagram with $g$ on the horizontal axis and $\eta$ on the vertical axis. }
\end{minipage}
\end{figure}
\section{Dual transformation from the AXY model to the gauged DAXY model by Villain approximation in a 2+1d system \label{sec4}}
In this section, we perform the dual transformation between the $A\!X\!Y$ model and the $D\!A\!X\!Y$ model from the Villain approximation.\cite{ref25}-\cite{ref27}.First, we apply the Villain approximation to ${Z'}\!_{A\!X\!Y}$ introduced in Eq.(\ref{eq15}) as follows:
\begin{align}\label{eq32}
\scalebox{0.88}{$\displaystyle
Z\!_{Q\!V}\!\equiv\!R\!_{Q\!V}\!\!\!\int\!\!\!{D\theta }\!\sum\limits_{\left\{ n \right\}}\!\exp\!\sum\limits_{x,\tau }\Bigl\{\frac{-( {E'}_J^0 )_v}{2}\Bigl( \nabla_{\tau }\theta -2\pi {n_0} \Bigr)^2\!
\Bigr.
$}\nonumber\\
\scalebox{0.88}{$\displaystyle
\Bigl.  
\!+\!\frac{-\left({E'}_J \right)_v}{2}\sum\limits_{j=1}^2{\Bigl( \nabla_j\theta-2\pi {n_j} \Bigr)}^2 
\Bigr\}
 $},
\end{align}
where \scalebox{0.9}{$Z\!_{Q\!V}$} is the Villain approximation of the partition function \scalebox{0.9}{${Z'}\!_{A\!X\!Y}$}, and \scalebox{0.9}{$R\!_{Q\!V}\!\!\equiv\!\!{[R_v{\left( E_J \right)^2}R_v({E'}_J^0)]^{MM_{\tau }}}$} is the Villain’s normalization parameter,, where \scalebox{0.9}{${R_v}\!\left( E \right)\!\!\equiv\!\!\sqrt{2\pi {(E)_v}}{I_0}\!\left( E \right)$}, and \scalebox{0.9}{$(E)_v\!\!\equiv\!\!{\left\{ -2\ln[I_0(E)/I_1(E)]\right\}}^{-1}$}. \scalebox{0.9}{${I_0}\!\left( E \right)$} and \scalebox{0.9}{${I_1}\!\left( E \right)$} represents the modified Bessel functions of order zero and order one, respectively. The summation symbols \scalebox{0.9}{$\sum\limits_{\left\{ n \right\}}\!\!\equiv\!\!\prod\limits_{x,\tau}\!\prod\limits_{j=0}^2\sum\limits_{n_j\left( x,\tau  \right)=-\infty }^{\infty }$}  are used for the integer fields \scalebox{0.9}{$n_j\left( x,\tau  \right)$}. The partition function in Eq.(\ref{eq32}) is equivalent to the Euclidean version of  the quantum vortex dynamics in a film of superfluid helium introduced by Kleinert\cite{ref25,ref28}. For Eq.(\ref{eq32}) the following identities associated with the Jacobi theta function are used: 
\begin{align}\label{eq33}
\scalebox{0.84}{$\displaystyle 
\sum\limits_{n=-\infty }^{\infty }\!\!\!\exp\!\left\{ \frac{-E}{2}\left( \theta -2\pi n \right)^2\right\}\!=\!\!\!\sum\limits_{l=-\infty }^{\infty }\!\!\!{\frac{1}{\sqrt{2\pi E}}\!\exp\!\left( \frac{-b^2}{2E}+ib\theta  \right)}
$},
\end{align}
As a result, Eq.(\ref{eq32}) can be rewritten as follows:
\begin{align}\label{eq34}
\scalebox{0.85}{$\displaystyle 
Z\!_{QV}\!=\!C\!_{QV}\!\!\sum\limits_{\left\{ b \right\}}{\delta_{{\nabla_j}b_j,0}}\!\exp\!\sum\limits_{x,\tau }\!{\left[ \frac{-b_0^2\left( x \right)}{2{{\left( {E'}\!_J^0 \right)}_v}}+\sum\limits_{j=1}^{2}\frac{-b_j^2\left( x \right)}{2{{\left( {E'}\!_J \right)}_v}}\right]}
$},
\end{align}
where ${C_{QV}}$ is a normalization parameter defined by \scalebox{0.9}{${{[{I_0}{{\left( {E'}_J \right)}^2}{I_0}({E'}_J^0)]}^{MM_{\tau }}}$}, and ${b_i}\left( x \right)$ represents auxiliary magnetic fields with integer values. The integer dual vector potentials ${\tilde{a}_i}\!\left( x \right)$ ($i = 0,1,2$) is introduced as follows \cite{ref26}:
\begin{align}\label{eq35}
\scalebox{0.98}{$\displaystyle 
b_i\!\left( x \right)={{\varepsilon }_{ijl}}{\nabla _j}{\tilde{a}_l}\!\left( x \right)={\left( \nabla \times \mathbf{\tilde{a}} \right)_i}\!\left( x \right)
$},          
\end{align}
where ${\varepsilon _{ijl}}$ is the Levi–Civita symbol of three dimensions. By using the dual transformations of Eq.(\ref{eq35}), the following Eq.(\ref{eq36}) can be obtained:
\begin{align}\label{eq36}
\scalebox{0.73}{$\displaystyle
Z\!_{Q\!V}\!\equiv\!{C\!_{Q\!V}}\!\!\sum\limits_{\left\{ {\tilde{a}} \right\}}{\delta_{{\nabla_j}{\varepsilon_{ijl}}{\nabla_j}{\tilde{a}_l},0}}\!\exp\!\sum\limits_{x}\left[\frac{-{{\left( \nabla \times \mathbf{\tilde{a}} \right)}_0}^2}{2{{\left( {E'}_J^0 \right)}_v}}\!+\!\sum\limits_{j=1}^2\frac{-{\left( \nabla \times \mathbf{\tilde{a}} \right)_j}^2}{2{{\left( {{E'}_J} \right)}_v}} 
\right]
 $},
\end{align}
Substituting Poisson's formula in Eq.(\ref{eq37}) to (\ref{eq36}):
\begin{align}\label{eq37}
\scalebox{0.8}{$\displaystyle
\sum\limits_{\left( {\tilde{a}_j} \right)\!=\!-\infty }^{\infty }\!\!\!\!\!\!f\left( {\tilde{a}_j} \right)\!=\!\!\!\!\int\limits_{-\infty }^{\infty }\!\!\!D{{\tilde{\alpha }'}_j}f\!( \tilde{\alpha'}_j )\!\!\!\!\!\!\!\sum\limits_{\left( {\tilde{l}_j} \right)=-\infty }^{\infty }\!\!\!\!\!\!{\delta_{{\nabla _j}{\tilde{l}_j},0}\!\exp\!\sum\limits_{x}\!\!{\left( i2\pi \sum\limits_{j=0}^2{{\tilde{l}_j}{\tilde{\alpha'}_j}} \right)}}
$},
\end{align}
then Eq..(\ref{eq38}) is as follows:
\begin{align}\label{eq38}
\scalebox{0.8}{$\displaystyle
Z\!_{Q\!V}\!=\!C\!_{Q\!V}\!\!\int\!\!{D{\tilde{\alpha'}_i}}\sum\limits_{\left\{ {\tilde{l}_i} \right\}}{{\delta _{{\nabla_j}{\tilde{l}_j},0}}}\!\exp\!\sum\limits_{x}\left\{ \frac{-\left( \nabla \times \mathbf{\tilde{\alpha'}} \right)_0^2\left( x \right)}{2{{\left( {E'}_J^0 \right)}_v}}
\right.
$}\nonumber\\
\scalebox{0.8}{$\displaystyle
\left.  
+\sum\limits_{j=1}^2\frac{-\left( \nabla \times \mathbf{\tilde{\alpha'}} \right)_j^2\left( x \right)}{2{{\left( {{E'}_J} \right)}_v}}
\!+i2\pi\!\sum\limits_{j=0}^2{\tilde{l}_j\left( x \right)\tilde{\alpha'}_j\left( x \right)} 
\right\}
 $},
\end{align}
Poisson's formula in Eq.(\ref{eq37}) converts the integer-valued vector potentials $\tilde{a}_i$ to the continuous-valued vector potentials ${\tilde{\alpha }'}_i$. Following the quantum vortex dynamics \cite{ref25,ref28}, the Euclidean Lagrangian density of ${\tilde{\alpha }'}_i$ is as follows:
\begin{align}\label{eq39}
\scalebox{0.85}{$\displaystyle
L_{QV}\!\!\left( x \right)\!=\!\frac{\tilde{\beta }_0^2\left( x \right)}{2{{\left( {E'}_J^0 \right)}_v}}+\sum\limits_{i=1}^2\frac{\tilde{\beta }_i^2\left( x \right)}{2{{\left( {E'}_J\right)}_v}}-i2\pi \sum\limits_{j=0}^2\!{{\tilde{l}_j}\!\left( x \right){{\tilde{\alpha' }}_j}\!\left( x \right)}
$},
\end{align}
where, ${\tilde{\beta }_i}$($i = 1,2$) and ${\tilde{\beta }_0}$ can be considered as a dual electric field and a dual magnetic field in a  $2 +1d$ dual electromagnetic field, respectively, and are defined as follows:
\begin{align}\label{eq40}
\scalebox{0.9}{$\displaystyle
{\tilde{\beta }_0}\left( x \right)\equiv{\nabla_1}{\tilde{\alpha'}_2}\left( x \right)-{{\nabla }_2}{\tilde{\alpha'}_1}\left( x \right)
$},\nonumber\\
\scalebox{0.9}{$\displaystyle
{\tilde{\beta }_1}\left( x \right)\equiv{\nabla _2}{\tilde{\alpha'}_0}\left( x \right)-{{\nabla }_0}{\tilde{\alpha'}_2}\left( x \right)
$},\nonumber\\
\scalebox{0.9}{$\displaystyle
{\tilde{\beta }_2}\left( x \right)\equiv{\nabla _0}{\tilde{\alpha'}_1}\left( x \right)-{{\nabla }_1}{\tilde{\alpha'}_0}\left( x \right)
$},
\end{align}
If the 1,2 components ${\tilde{e}_1}$ and ${\tilde{e}_2}$ of the dual electric field are set as ${\tilde{e}_1}\equiv {\tilde{\beta }_2}$, and ${\tilde{e}_2}\equiv -{\tilde{\beta }_1}$, respectively, the dual Maxwell's equations from Lagrangian of Eq.(\ref{eq39}) are as follows:
\begin{align}\label{eq41}
\scalebox{0.98}{$\displaystyle
\frac{1}{\left( {E'}\!_J \right)_v}\left[ {\nabla_1}{\tilde{e}_1}\left( x \right)+{\nabla_2}{\tilde{e}_2}\left( x \right) \right]=i2\pi {\tilde{l}_0}\left( x \right)
$},\nonumber\\
\scalebox{0.98}{$\displaystyle
 \frac{1}{\left( {E'}\!_J^0 \right)_v}{\nabla_2}{\tilde{\beta }_0}\left( x \right)-\frac{1}{\left( {E'}\!_J \right)_v}{\nabla_0}{\tilde{e}_1}\left( x \right)=i2\pi {\tilde{l}_1}\left( x \right)
$},\nonumber\\
\scalebox{0.98}{$\displaystyle
-\frac{1}{\left( {E'}\!_J \right)_v}{\nabla_0}{\tilde{e}_2}\left( x \right)-\frac{1}{\left( {E'}\!_J^0 \right)_v}{\nabla_1}{\tilde{\beta }_0}\left( x \right)=i2\pi{\tilde{l}_2}\left( x \right)
$},
\end{align}
For Eq.(\ref{eq38}), in order to integrate out for continuous-valued vector potentials $\tilde{\alpha'}_i$,  the partition function when the axial gauge-fixing condition $\tilde{\alpha'}_0=0$ is imposed is as follows:  
\begin{align}\label{eq42}
\scalebox{0.82}{$\displaystyle
Z\!_{Q\!V}\!=\!C\!_{Q\!V}\!\!\int{D{\tilde{\alpha'}_1}}\int{D{\tilde{\alpha'}_2}}\sum\limits_{\left\{ \tilde{l}_i \right\}}{{\delta }_{\nabla_j\tilde{l}_j,0}}\exp\Bigl[\frac{-1}{2{\left( {E'}_J^0 \right)_v}}
\Bigr.
$}\nonumber\\
\scalebox{0.82}{$\displaystyle
\Bigl.  
\times\sum\limits_{x,x'}{\tilde{\alpha' }_i}^{\bot }\left( x \right)D_{ij}^{\bot }\left( x,{x'} \right){\tilde{\alpha' }_j}^{\bot }\left( x' \right)\!+\!i2\pi\!\sum\limits_{j=0}^2{{\tilde{l}_j}^{\bot }\!\left( x \right){\tilde{\alpha' }_i}^{\bot }\left( x \right)} 
\Bigr]
 $},\nonumber\\
\scalebox{0.82}{$\displaystyle 
 D_{ij}^{\bot }\left( x,x' \right)\equiv -{{\delta }_{ij}}^{\bot }{{g}^{ab}}{{\bar{\nabla }}_{a}}{{\nabla }_{b}}+{{\bar{\nabla }}_{i}}^{\bot }{{\nabla }_{j}}^{\bot }
 $},
\end{align}
where the orthogonal symbol $\bot$ as a superscript indicates that only the components ${{\tilde{\alpha }'}_1}$ and ${{\tilde{\alpha }'}_2}$ exist,  and the metric $g^{ab}$ is defined as follows:
\begin{align}\label{eq43}
\scalebox{0.80}{$\displaystyle
g^{ab}\equiv \begin{pmatrix}
   \gamma  & 0 & 0  \\[-12pt]
   0 & 1 & 0  \\[-12pt]
   0 & 0 & 1  \\
\end{pmatrix},\;\;\;
\gamma \equiv \frac{\left( E_J^0 \right)_v}{\left( E_J \right)_v} 
$},
\end{align}
where $\gamma$ is an anisotropic parameter in the $J\!J$ model. Integrating over the continuous-valued gauge fields ${{\tilde{\alpha }'}_1}$ and  ${{\tilde{\alpha }'}_2}$ for Eq.(\ref{eq42}) yields following the partition function:
\begin{align}\label{eq44}
\scalebox{0.78}{$\displaystyle
Z\!_{Q\!V}\!=\!C'\!_{Q\!V}\!\!\sum\limits_{\left\{ {l_i} \right\}}{{\delta_{{\tilde{l}_i}\left( x \right),0}}}\!\exp\!\sum\limits_{x,x'}\!\Bigl[-2{{\pi }^2}{{( {E'}_J^0 )}_v}{{\tilde{l}}_j}\left( x \right){{\cal{V}}_0}\left( x-x' \right){\tilde{l}_j}\left( x' \right)
\Bigr.
$}\nonumber\\
\scalebox{0.78}{$\displaystyle
\Bigl.  
-2{{\pi }^2}{{( {E'}_J)}_v}{\tilde{l}_0}\left( x \right){{\cal{V}}_0}\left( x-x' \right){\tilde{l}_0}\left( x' \right)
\Bigr]
 $},
\end{align}
where \scalebox{0.85}{${C'_{QV}}\equiv {C_{QV}}{{\left[ \det \left( -{{\eta }^{ab}}{\bar{\nabla }_a}{\nabla_b} \right) \right]}^{\frac{-1}{2}}}{{\left[ \det \left( -{\bar{\nabla }_0}{\nabla_0} \right) \right]}^{\frac{-1}{2}}}$}, and the anisotropic massless lattice potential (or lattice Green's function)\cite{ref25} ${{\cal{V}}_0}\left( x \right)$ is defined as:
\begin{align}\label{eq45}
\scalebox{0.85}{$\displaystyle
{{\cal{V}}_0}\left( x \right)\equiv \frac{-1}{g^{ab}\bar\nabla_a\nabla_b}\left( x\right)
 $},\quad
\end{align}
Now, from this lattice potential, the "split lattice potential" \scalebox{0.85}{${{\cal{V'}}_0}\left( x \right)$} obtained by dividing the "core lattice potential"\scalebox{0.85}{${\cal{V}}_0\left( 0 \right)$} and "split difference operator"${{\nabla'}_i}$ are introduced as follows:
\begin{align}\label{eq46}
\scalebox{0.85}{$\displaystyle
{{\cal{V'}}_0}\left( x \right)\equiv \frac{-1}{g^{ab}\bar{\nabla'}_a{\nabla'}_b}\left( x \right),
$}\quad\quad\nonumber\\ 
\scalebox{0.85}{$\displaystyle
{{\nabla' }_i}\equiv \frac{\nabla_i}{\sqrt{1-{{\cal{V}}_0}\left( 0 \right)\left( -{g^{ab}}{\bar{\nabla}_a}{\nabla_b} \right)}}
$}\quad
\end{align}
Thus, we have the following identity:
\begin{align}\label{eq47}
\scalebox{0.64}{$\displaystyle
\exp\!\sum\limits_{x,x'}\!{\left[\!-2{{\pi }^2}{{\left( {E'}_J^0 \right)}_v}\!\!{\tilde{l}_j}\left( x \right)\!{{\cal{V'}}_0}\!\left( x-x' \right)\!{\tilde{l}_j}\!\left( {x'} \right)\!-\!2{{\pi }^2}{{\left({E'}_J \right)}_v}{\tilde{l}_0}\left( x \right){{\cal{V'}}_0}\left( x-x' \right){\tilde{l}_0}\left( {x'} \right) \right]}
$}\nonumber\\ 
\scalebox{0.64}{$\displaystyle
\!=\!C'\!\!\!\int{D\tilde{\alpha'}_i}\exp\!\sum\limits_{x}{\left[ \frac{-\left( {\nabla'}\times \mathbf{\tilde{\alpha'}}\right)_0^2}{2\left({E'}_J^0 \right)_v}+\sum\limits_{j=1}^2\frac{-{\left( {\nabla' }\times \mathbf{\tilde{\alpha' }} \right)_j^2}}{2\left({E'}_J\right)_v}\right]}
$},\quad
\end{align}
where \scalebox{0.75}{$C'\!\equiv\!{{\left[ \det \left( -\tilde{\bar{\nabla' }}_l\tilde{\nabla' }_l\right) \right]}^{\frac{1}{2}}}{{\left[ \det \left( -\tilde{\bar{\nabla'}}_0{\tilde\nabla'}_0\right) \right]}^{\frac{1}{2}}}$}. Substituting Eqs.(\ref{eq46}) and (\ref{eq47}) in Eq.(\ref{eq44}) gives:
\begin{align}\label{eq48}
\scalebox{0.7}{$\displaystyle
Z_{QV}\!=\!C''\!_{QV}\!\!\!\int{D{\tilde\alpha'_i}}\!\exp\!\sum\limits_{x}\!\!{\left[ \frac{-1}{2{{\left( {E'}_J^0 \right)}_v}}f^{ab}{{\left( \nabla'\times \mathbf{\tilde{\alpha' }} \right)}_a}\!{{\left( {\nabla' }\times \mathbf{\tilde{\alpha' }} \right)}_b}\!\right]}\!\!\sum\limits_{\left\{ {{\tilde{l}}_i} \right\}}{{\delta_{{\tilde{l}}_i,0}}}
$}\nonumber\\
\scalebox{0.7}{$\displaystyle
\!\times\!\exp\!\sum\limits_{x}{\left[ -2{{\pi }^2}{{\cal{V}}_0}\left( 0 \right){{\left( {E'}_J^0 \right)}_v}{{\tilde{l}}_j}^2-2{{\pi }^2}{{\cal{V}}_0}\left( 0 \right){{\Bigl( {E'}_J \Bigr)}_v}{\tilde{l}_0}^2+i2\pi \sum\limits_{j=0}^2{{{\tilde{l}}_j}{{{\tilde{\alpha' }}}_j}} \right]}                  
 $},
\end{align}
where \scalebox{0.75}{${C''_{QV}}\!\equiv\!{C'_{QV}}{{\left[ \det \left( -{\tilde{\bar{\nabla'}}_l}{\tilde{\nabla'}_l} \right) \right]}^{\frac{1}{2}}}{{\left[ \det \left( -{\tilde{\bar{\nabla'}}_0}{\tilde{\nabla'}\!_0} \right) \right]}^{\frac{1}{2}}}$}, and the metric \scalebox{0.78}{${f^{ab}}$} is defined as::
\begin{align}\label{eq49}
\scalebox{0.90}{$\displaystyle
f^{ab}\equiv
\begin{pmatrix} 
1 & 0 & 0 \\[-12pt]
0 & \gamma & 0 \\[-12pt]
0 & 0 & \gamma
\end{pmatrix}
$}.
\end{align}
Furthermore, from Eqs.(\ref{eq15}), (\ref{eq32}), and (\ref{eq34}), the following identity can be obtained:
\begin{align}\label{eq50}
\scalebox{0.62}{$\displaystyle
\sum\limits_{\left\{ \tilde{l}_i \right\}}\!{\delta_{\tilde{l}_i,0}}\!\exp\!\!\sum\limits_{x}\!\left[-2{\pi }^2{{\cal{V}}_0}\left( 0 \right)\!{\left( {E'}_J^0 \right)}_v\!{\tilde{l}_j}^2-2{\pi }^2{{\cal{V}}_0}\left( 0 \right)\!{\Bigl( {E'}\!_J \Bigr)}_v\!{\tilde{l}_0}^2+i2\pi \sum\limits_{j=0}^{2}{{\tilde{l}_j}\tilde{\alpha'}_j} 
 \right]
$}\nonumber\\ 
\scalebox{0.62}{$\displaystyle
\approx\!\!\!{\int\!\!D\tilde{\theta }}\exp\!\left( -1 \right)\!\sum\limits_{x}\!{\left\{ \frac{\left[ 1-\cos \left( {\nabla_0}\tilde{\theta }-2\pi {\tilde{\alpha' }_0} \right)\right]\!\!}{4{\pi }^2{{\cal{V}}_0}\left( 0 \right){\Bigl( {{E'}_J} \Bigr)}_v}+\frac{\sum\limits_{j=1}^2{\left[ 1-\cos \left( {\nabla }_j\tilde{\theta }-2\pi {\tilde{\alpha'}_j} \right) \right]\!\!} }{4{{\pi }^2}{{\cal{V}}_0}\left( 0 \right){{\Bigl({E'}_J^0 \Bigr)}_v}}\right\}}
$},
\end{align}
As an analogy to the relational expression between the $J\!J$ energy and the $Q\!P\!S\!J$ energy in Eq.(\ref{eq12}),the following relational expression is obtained from Eq.(\ref{eq50}).
\begin{align}\label{eq51}
\scalebox{0.85}{$\displaystyle
{{\left({E'}\!_S^0 \right)}_v}\!\!\equiv\!\frac{1}{4{{\pi }^2}{{\cal{V}}_0}\!\left( 0 \right)\!{{\left( {{E'}\!_J} \right)}_v}},\;\;{{\left( {{E'}\!_S} \right)}_v}\!\!\equiv\!\frac{1}{4{{\pi }^2}{{\cal{V}}_0}\!\left( 0 \right)\!{{\left( {E'}\!_J^0 \right)}_v}}, 
$}
\end{align}
Eqs.(\ref{eq12}) and (\ref{eq51}) differ only in the coefficients of the core lattice potential ${V_0}\left( 0 \right)$, therefore, Eq.(\ref{eq51}) is a reasonable result. Substituting Eqs.(\ref{eq50}) and (\ref{eq51}) in Eq.(\ref{eq48}) gives: 
\begin{align}\label{eq52}
\scalebox{0.66}{$\displaystyle
Z_{QV}\!=\!{{C''}\!\!_{QV}}\!\!\int{\!\!D{\tilde{\alpha'}_i}}\exp\!\sum\limits_{x}{\left[ \frac{-f^{ab}}{2{{\left( {E'}_J^0 \right)}_v}}{{\left( {\nabla'}\times \mathbf{\tilde{\alpha'}} \right)}_a}{{\left( {\nabla }'\times \mathbf{\tilde{\alpha'}} \right)}_b}\right]}\!\!\!\int{\!D\tilde{\theta }}\exp\!\left( -1 \right)\!\!
$}\nonumber\\ 
\scalebox{0.66}{$\displaystyle
\times\!\sum\limits_{x}{\left\{ \!\!{( {E'}_S^0)}_v\!\!\left[ 1\!-\!\cos \left( {\nabla_0}\tilde{\theta }-2\pi {\tilde{\alpha'}_0} \right) \right]\!+\!{{({E'}\!_S)}_v}\!\!\sum\limits_{j=1}^2{\left[ 1\!-\!\cos\!\left({\nabla }_j\tilde{\theta }-2\pi {{\tilde{\alpha'}}_j} \right) \right]}\right\}} 
$},
\end{align}
Eq.(\ref{eq52}) represents the gauge-coupled $Q\!P\!S\!J$ partition function by the dual gauge field $\tilde{\alpha'}_i$.
For the coefficient ${{\left( {E'}\!_J^0 \right)}\!_v}$ of the dual gauge field energy, use the relationships in Eqs.(\ref{eq51}) and (\ref{eq7}), and next, when these non-dimensional energy constants have been converted to the original energy dimension, Eq. (and next, when these non-dimensional energy constants have been converted to the original energy dimension, Eq. (52) will be as follows:) will be as follows: 
\begin{align}\label{eq53}
\scalebox{0.62}{$\displaystyle
Z\!_{QV}\!=\!{C''}\!_{QV}\!\!\!\int\!\!{D{\tilde{\alpha' }\!_i}}\exp\!\!\sum\limits_{\tau,\bm{x}}\!\left(\! \frac{-a_0}{\hbar }\! \right)\!\!\!{\left[ {\cal{V}}_0\!\!\left( 0 \right)\!\frac{{\left( 2e \right)}^2f^{ab}}{2C}{{\left( {\nabla }'\!\times\!\mathbf{\tilde{\alpha'}} \right)}_a}\!{{\left( {\nabla'}\!\times\!\mathbf{\tilde{\alpha' }} \right)}_b} \right]}\!\!\int\!\!D\tilde{\theta }
$}\nonumber\\
\scalebox{0.62}{$\displaystyle
\times\!\!\exp\!\!\left(\!\frac{-a_0}{\hbar }\!\right)\!\!\sum\limits_{\tau,\bm{x} }{\!\left\{\! (E_S^0)_v\!\!\left[ \!1\!-\!\cos\!\left( {{\nabla }_0}\tilde{\theta }\!-\!2\pi {{\tilde{\alpha' }}_0} \right)\!\right]\!+\!(E_S)_v\!\!\sum\limits_{j=1}^2\!{\left[ \!1\!-\!\cos\!\left( {{\nabla }_j}\tilde{\theta }\!-\!2\pi {{\tilde{\alpha' }}_j} \right)\!\right]}\!\right\}}
$},
\end{align}
Furthermore, by scaling $2e$ to the non-dimensional dual gauge field $\tilde{\alpha'}\!_i$, we introduce a new dual gauge field $\tilde{\alpha }_i$  as follows: 
\begin{align}\label{eq54}
\scalebox{0.95}{$\displaystyle
2e{\tilde{\alpha'}_i}\left( x \right)\equiv {\tilde{\alpha }_i}\left( x \right)
$},
\end{align}
By the transformation of Eq.(\ref{eq54}), Eq.(\ref{eq53}) is transformed as follows:
\begin{align}\label{eq55}
\scalebox{0.70}{$\displaystyle
Z\!_{QV}\!=\!{C''}\!_{QV}\!\!\!\int\!\!{D{\tilde{\alpha}\!_i}}\exp\!\!\sum\limits_{\tau,\bm{x}}\!\!\left(\! \frac{-a_0}{\hbar }\! \right)\!\!\!{\left[ \frac{f^{ab}}{2{\tilde{\mu'' }}}{{\left( {\nabla}\times \mathbf{\tilde{\alpha}} \right)}_a}\!{{\left( {\nabla}\times \mathbf{\tilde{\alpha}} \right)}_b} \right]}\!\!\int\!\!D\tilde{\theta }\!\exp\!\left(\!\frac{-a_0}{\hbar }\!\right)
$}\nonumber\\
\scalebox{0.70}{$\displaystyle
\times\!\sum\limits_{\tau,\bm{x}}\!\left\{ E_S^0\!\left[ 1\!-\!\cos\!\left( {\nabla }_0\tilde{\theta }-q_m {\tilde{\alpha}_0} \right) \right]\!+\!{E_S}\!\sum\limits_{j=1}^2{\left[ 1\!-\!\cos\!\left( {\nabla }_j\tilde{\theta }-q_m{\tilde{\alpha}_j} \right) \right]}\!\right\}
$},
\end{align}
where \scalebox{0.90}{$\tilde{\mu'' }\!\equiv\!C/V_0\!\left( 0 \right)$}, and \scalebox{0.90}{$q_m\!\equiv\!2\pi\!/2e\!=\!\Phi _0\!/\hbar$} represents a unit magnetic charge. Eq.(\ref{eq55}) shows that the pure \scalebox{0.95}{$A\!X\!Y$} model (\scalebox{0.95}{$J\!J$} model without gauge coupling) has been dual transformed to the gauged \scalebox{0.95}{$D\!A\!X\!Y$} model (gauged \scalebox{0.95}{$Q\!P\!S\!J$} model) by the Villain approximation in the $2+1 d$ system. In other words, the gauge \scalebox{0.95}{$Q\!P\!S\!J$} model is a “frozen lattice dual superconductor”\cite{ref25,ref29}-\cite{ref31}, for the $J\!J$ model without gauge coupling.
\section{Dual transformation from the DAXY model to the gauged AXY model by Villain approximation in a 2 + 1 d system \label{sec5}}
In this section, contrary to the previous section, we show the dual transformation from the $D\!A\!X\!Y$ model to the $A\!X\!Y$ model by the Villain approximation. First, apply the Villain approximation to ${Z'}_{D\!A\!X\!Y}$ introduced in Eq.(\ref{eq16}) as follows:
\begin{align}\label{eq56}
\scalebox{0.63}{$\displaystyle
Z_{Q\!D\!V}\!\!\equiv\!\!{R_{Q\!D\!V}}\!\!\!\int\!\!\!D\tilde{\theta }\sum\limits_{\left\{\tilde{n}\right\}}\exp\!\sum_{x}\!\!\left[\frac{\!\!-\!{\left(\!{E'}_S^0 \!\right)_v}}{2}{{\Bigl(\!{\nabla }_{\tau }\tilde{\theta }\!-\!2\pi{\tilde{n}_0}\!\Bigr)}\!^2}\!\!
+\!\!\frac{-{{\Bigl(\! {E'_S} \!\Bigr)}_v}}{2}\!\!\sum\limits_{j=1}^{2}\!{{\Bigl(\! {{\nabla }_j}\tilde{\theta }\!-2\pi{\tilde{n}_j}\!\Bigr)}\!^2} 
\!\right] $},
\end{align}

where \scalebox{0.88}{$R_{D\!Q\!V}\!\equiv\!{{[R_v\!\left( {E'}\!_S \right)}^2{R_v}\!({E'}\!_S^0)]}^{M{M_\tau}}$}, and $Z\!_{Q\!D\!V}$ represents the Villain approximations of the partition function ${Z'}\!_{D\!A\!X\!Y}$. Using the Jacobi theta function of Eq.(\ref{eq33}), Eq.(\ref{eq56}) can be rewritten as follows:
\begin{align}\label{eq57}
\scalebox{0.85}{$\displaystyle 
Z\!_{Q\!D\!V}\!=\!C\!_{Q\!D\!V}\!\!\sum\limits_{\left\{ \tilde{b} \right\}}{\delta_{{\nabla_j}\tilde{b}_j,0}}\exp\sum\limits_{x}\!{\left[ \frac{-\tilde{b}_0^2\left( x \right)}{2{{\left( {E'}\!_S^0 \right)}_v}}+\sum\limits_{j=1}^2\frac{-\tilde{b}_j^2\left( x \right)}{2{{\left( {E'}\!_S \right)}_v}} \right]}
$},
\end{align}
where \scalebox{0.88}{${C_{Q\!D\!V}}\!\equiv\!{{[{I_0}{{\left({E'}_J \right)}^2}{I_0}({E'}_J^0)]}^{M{M_\tau}}}$}, and ${\tilde{b}_i}\left( x \right)$ represents auxiliary dual magnetic fields with integer values. Integer vector potentials ${a_i}\left( x \right)$($i \!= \!0,1,2$) are introduced as follows : 
\begin{align}\label{eq58}
\scalebox{0.98}{$\displaystyle 
\tilde{b}_i\!\left( x \right)\!=\!{{\varepsilon }_{ijl}}{\nabla _j}{a_l}\!\left( x \right)={\left( \nabla \times \mathbf{a} \right)_i}\!\left( x \right)
$},          
\end{align}
By using the dual transformations of Eq.(\ref{eq58}), the following Eq.(\ref{eq59}) is obtained:
\begin{align}\label{eq59}
\scalebox{0.76}{$\displaystyle
Z\!_{Q\!D\!V}\!=\!C\!_{Q\!D\!V}\!\!\sum\limits_{\left\{ {a} \right\}}\!{\delta_{{\nabla\!_j}{\varepsilon\!_{ijl}}{\nabla\!_j}{a_l},0}}\!\exp\!\sum\limits_{x}\!\!\left[\frac{-{{\left( \nabla\!\times\!\mathbf{a} \right)}_0}^2}{2{{\left( {E'}_S^0 \right)}_v}}\!+\!\sum\limits_{j=1}^2\!\frac{-{\left( \nabla\!\times\!\mathbf{a} \right)_j}^2}{2{{\left( {E'}_S \right)}_v}} \right]
$},
\end{align}
Using Poisson's formula in the following Eq.(\ref{eq37}) for Eq.(\ref{eq59}):
\begin{align}\label{eq60}
\scalebox{0.8}{$\displaystyle
Z\!_{Q\!D\!V}\!=\!C\!_{Q\!D\!V}\!\!\int\!\!{D{\alpha'}\!_i}\!\sum\limits_{\left\{ {l_i} \right\}}{{\delta _{{\nabla_j}{l_j},0}}}\!\exp\!\sum\limits_{x}\Bigl[\frac{-1}{2{{\left({E'}_S\right)}_v}}\!\sum\limits_{j=1}^2\!{\left( \nabla \times \mathbf{\alpha'}\right)_{j}^{2}\!}
\Bigr.
$}\nonumber\\
\scalebox{0.8}{$\displaystyle
\Bigl.  
+\frac{-1}{2{{\left( {E'}_S^0 \right)}_v}}\left( \nabla \times \mathbf{\alpha'}\right)_{0}^{2}+i2\pi\!\sum\limits_{j=0}^2{{l_j}{{\alpha'}_j}} 
\Bigr]
 $},
\end{align}
The Euclidean Lagrangian density of ${\alpha'}\!_j$ is as follows: 
\begin{align}\label{eq61}
\scalebox{0.75}{$\displaystyle
L\!_{Q\!D\!V}\!\!\left( x \right)\!\!=\!\!\frac{1}{2{{\left( {E'}\!_S\right)}_v}}\!\!\sum\limits_{i=1}^2{{\beta }_i^2\left( x \right)}\!+\!\frac{1}{2{{\left( {E'}\!_S^0 \right)}_v}}{\beta }_{0}^2\left( x \right)-i2\pi \sum\limits_{j=0}^2\!{{l_j}\!\left( x \right){{\alpha'}_j}\!\left( x \right)}
$},
\end{align}
 where,${\beta_i}$($i=1,2$) and ${\beta_0}$ can be considered as an electric field and a magnetic field in a $2+1d$ electromagnetic field, respectively, and are defined as follows:
\begin{align}\label{eq62}
\scalebox{0.9}{$\displaystyle
{\beta }_0\left( x \right)\equiv{\nabla_1}{{\alpha'}_2}\left( x \right)-{{\nabla }_2}{{\alpha'}_1}\left( x \right)
$},\nonumber\\
\scalebox{0.9}{$\displaystyle
{\beta }_1\left( x \right)\equiv{\nabla _2}{{\alpha'}_0}\left( x \right)-{{\nabla }_0}{{\alpha'}_2}\left( x \right)
$},\nonumber\\
\scalebox{0.9}{$\displaystyle
{\beta }_2\left( x \right)\equiv{\nabla _0}{{\alpha'}_1}\left( x \right)-{{\nabla }_1}{{\alpha'}_0}\left( x \right)
$},
\end{align}
If the 1,2 components ${{e}_{1}}$ and ${{e}_{2}}$ of the electric field are set as ${{e}_{1}}\equiv {{\beta }_{2}}$, and ${{e}_{2}}\equiv -{{\beta }_{1}}$, respectively, the Maxwell's equations from the Lagrangian of Eq.(\ref{eq61}) are as follows:
\begin{align}\label{eq63}
\scalebox{0.9}{$\displaystyle
\frac{1}{\left( {E'}\!_S \right)_v}\left[ {\nabla_1}{e_1}\left( x \right)+{\nabla_2}{e_2}\left( x \right) \right]=i2\pi {l_0}\left( x \right)
$},\nonumber\\
\scalebox{0.9}{$\displaystyle
 \frac{1}{\left( {E'}\!_S^0 \right)_v}{\nabla_2}{\beta_0}\left( x \right)-\frac{1}{\left( {E'}\!_S \right)_v}{\nabla_0}{e_1}\left( x \right)=i2\pi {l_1}\left( x \right)
$},\nonumber\\
\scalebox{0.9}{$\displaystyle
-\frac{1}{\left( {E'}\!_S \right)_v}{\nabla_0}{e_2}\left( x \right)-\frac{1}{\left( {E'}\!_S^0 \right)_v}{\nabla_1}{{\beta }_0}\left( x \right)=i2\pi{l_2}\left( x \right)
$},
\end{align}
Integrating over the continuous-valued gauge fields ${{\alpha'}\!_1}$ and ${{\alpha'}\!_2}$ for Eq.(\ref{eq60}) yields the following partition function:
\begin{align}\label{eq64}
\scalebox{0.78}{$\displaystyle
Z\!_{Q\!D\!V}\!=\!C'\!_{Q\!D\!V}\!\!\sum\limits_{\left\{ {l_i} \right\}}{{\delta_{{l_i}\left( x \right),0}}}\!\exp\!\sum\limits_{x,x'}\!\Bigl[\!-\!2{{\pi }^2}{{\left( {E'}_J^0 \right)}_v}\!{l_j}\left( x \right)\!{\tilde{\cal{V}}_0}\!\left( x-x' \right)\!{l_j}\left( x' \right)
\Bigr.
$}\nonumber\\
\scalebox{0.78}{$\displaystyle
\Bigl.  
\!-\!2{{\pi }^2}{{\Bigl( {E'}_J \Bigl)}_v}\!{l_0}\left( x \right)\!{\tilde{\cal{V}}_0}\!\left( x-x' \right)\!{l_0}\left( x' \right)
\Bigr]
 $},
\end{align}
where \scalebox{0.9}{${{C'}\!_{Q\!D\!V}}\equiv {C\!_{Q\!D\!V}}{{\left[ \det \left( -{{\eta }^{ab}}{{\bar{\nabla }}_a}{{\nabla }_b} \right) \right]}^{\frac{-1}{2}}}{{\left[ \det \left( -{{\bar{\nabla }}_0}{{\nabla }_0} \right) \right]}^{\frac{-1}{2}}}$}, and the anisotropic massless lattice potential (or lattice Green's function) $\tilde{V}_0\left(x\right)$ is defined as:
\begin{align}\label{eq65}
\scalebox{0.83}{$\displaystyle
{\tilde{\cal{V}}_0}\left( x \right)\equiv\frac{-1}{{{\tilde{g}}^{ab}}{\bar{\nabla }_a}{\nabla_b}}\left( x \right),
$}\nonumber\\  
\scalebox{0.83}{$\displaystyle      
{{\tilde{g}}^{ab}}\equiv \left( 
\begin{matrix}
   {\tilde{\gamma }} & 0 & 0  \\[-12pt]
   0 & 1 & 0  \\[-12pt]
   0 & 0 & 1  
\end{matrix} 
\right),\;\; \tilde{\gamma }\equiv \frac{{\left( E_S^0 \right)_v}}{\left( {E_S} \right)_v}
$},
\end{align}
From this lattice potential, we introduce a "split lattice potential" \scalebox{0.9}{$\tilde{\cal{V'}}_0\left( x \right)\equiv {\tilde{\cal{V}}_0}\left( x \right)\!-\!{{\tilde{\cal{V}}}_{0}}\left( 0 \right){{\delta }_{x,0}}$} which is obtained by dividing the "core lattice potential" \scalebox{0.9}{${\tilde{\cal{V}}}_0\left( 0 \right)$} and "split difference operator" \scalebox{0.9}{${\nabla' }_j$} , as follows:
\begin{align}
\scalebox{0.85}{$\displaystyle
{{\tilde{\cal{V'}}}_0}\left( x \right)\equiv \frac{-1}{\tilde{g}^{ab}\bar{\nabla'}_a{\nabla'}_b}\left( x \right),
$}\quad\quad\nonumber
\end{align}
\begin{align}
\scalebox{0.85}{$\displaystyle
{{\nabla' }_i}\equiv \frac{\nabla_i}{\sqrt{1-{{\tilde{\cal{V}}}_0}\left( 0 \right)\left( -{\tilde{g}^{ab}}{\bar{\nabla}_a}{\nabla_b} \right)}}
$}\quad\nonumber
\end{align}
Similar from Eq.(\ref{eq47}) to (\ref{eq51}) in the previous section, the partition function ${Z_{DQV}}$ can be written as follows:
\begin{align}\label{eq66}
\scalebox{0.65}{$\displaystyle
Z\!_{Q\!D\!V}\!=\!\!{{C''}\!\!_{Q\!D\!V}}\!\!\int{\!\!D{{\alpha'}_i}}\exp\!\sum\limits_{x}{\left[ \frac{-1}{2{{\left( {E'}_S^0 \right)}_v}}{f^{ab}}{{\left( {\nabla'}\times \mathbf{{\alpha'}} \right)}_a}{{\left( {\nabla }'\times \mathbf{{\alpha'}} \right)}_b}\right]}\!\!\!\int{\!D\tilde{\theta }}\exp\!\left( -1 \right)\!\!
$}\nonumber\\ 
\scalebox{0.65}{$\displaystyle
\!\times\!\sum\limits_{x}{\left\{ \!\!( {E'}_J^0 )_v\!\!\left[ 1\!-\!\cos \left( {\nabla_0}\tilde{\theta }-2\pi {{\alpha'}\!_0} \right) \right]\!+\!({E'}\!_J )_v\!\!\sum\limits_{j=1}^2{\left[ 1\!-\!\cos\!\left({\nabla }_j\tilde{\theta }-2\pi {{\alpha'}\!_j} \right) \right]}\right\}} 
$},
\end{align}
where \scalebox{0.75}{${{C''}\!_{Q\!D\!V}}\!\equiv\!{{C'}\!_{Q\!D\!V}}{{\left[ \det\!\left( -{{\tilde{\bar{\nabla'}}}_l}{{\tilde{\nabla' }}_l} \right) \right]}^{\frac{1}{2}}}{{\left[ \det\!\left( -{{\tilde{\bar{\nabla' }}}_0}{{\tilde{\nabla' }}_0} \right) \right]}^{\frac{1}{2}}}$}, and the metric \scalebox{0.8}{${{\tilde{f}}^{ab}}$} is defined as:
\begin{align}\label{eq67}
\scalebox{0.90}{$\displaystyle
{\tilde{f}}^{ab}\equiv
\begin{pmatrix} 
1 & 0 & 0 \\[-12pt]
0 & \tilde{\gamma} & 0 \\[-12pt]
0 & 0 & \tilde{\gamma}
\end{pmatrix}
$}
\end{align}
where \scalebox{0.8}{$( {E'}_J^0 )\!_v$} and \scalebox{0.8}{$( {E'}_J )\!_v$} are defined as follows:
\begin{align}\label{eq68}
\scalebox{0.82}{$\displaystyle
( {E'}_J^0 )_v\!\equiv\!\frac{1}{4{{\pi }^2}{\tilde{\cal{V}}_0}\left( 0 \right){( {E'}_S )}_v},  ( {E'}_J )_v\!\equiv\!\frac{1}{4{{\pi }^2}{{\tilde{\cal{V}}}_{0}}\left( 0 \right){{\left( {E'}_S^0 \right)}_v}}
$}
\end{align}
If \scalebox{0.85}{${{\cal{V}}_0}\left( 0 \right)\!\equiv\!{\tilde{\cal{V}}_0}\left( 0 \right)$} holds, Eqs.(\ref{eq68}) and (\ref{eq51}) are completely equivalent. Eq.(\ref{eq66}) represents the gauge-coupled $JJ$ partition function by the gauge field ${\alpha'}\!_j$. For the coefficient \scalebox{0.85}{${{( {E'}\!_S^0 )}_v}$} of the gauge field energy, use the relationship between Eqs.(\ref{eq68}) and (\ref{eq7}), and then, when these non-dimensional energy constants are converted to the original energy dimension, Eq. (\ref{eq68}) will be as follows:
\begin{align}\label{eq69}
\scalebox{0.64}{$\displaystyle
Z\!_{Q\!D\!V}\!=\!{C''}\!_{Q\!D\!V}\!\!\!\int\!\!{D{{\alpha' }\!_i}}\exp\!\!\sum\limits_{\tau,\bm{x}}\!\left(\! \frac{-a_0}{\hbar }\! \right)\!\!{\left[ {\tilde{\cal{V}}_0}\left( 0 \right)\!\!\left( 0 \right)\frac{{\left(\Phi_0 \right)}^2}{2L}{f^{ab}}{{\left( {\nabla }'\!\times\!\mathbf{{\alpha'}} \right)}_a}\!{{\left( {\nabla'}\!\times\!\mathbf{{\alpha' }} \right)}_b} \right]}\!\!\int\!\!D{\theta }
$}\;\;\nonumber\\
\scalebox{0.64}{$\displaystyle
\times\!\!\exp\!\!\left(\!\frac{-a_0}{\hbar }\!\right)\!\!\sum\limits_{\tau,\bm{x}}{\!\left\{ \!(E_J^0)_v\!\Bigl[ 1\!-\!\cos\!\left( {{\nabla }\!_0}{\theta }\!-\!2\pi {{\alpha' }\!\!_0} \right) \Bigr]\!+\!(E_J)_v\!\sum\limits_{j=1}^2\!{\Bigl[ 1\!-\!\cos\!\left( {{\nabla }\!\!_j}{\theta }\!-\!2\pi {{\alpha' }\!\!_j} \right) \Bigr]}\!\right\}}
$},
\end{align}
Furthermore, by scaling ${{\Phi }_0}$ to the non-dimensional gauge field ${{\alpha'}_i}$, we introduce a new gauge field ${{\alpha }_i}$ as follows: 
\begin{align}\label{eq70}
\scalebox{0.95}{$\displaystyle
 {\Phi_0}{{\alpha'}\!_i}\left( x \right)\!\equiv\!{\alpha_i}\left( x \right)
$},
\end{align}
Eq.(\ref{eq69}) is transformed as follows:         
\begin{align}\label{eq71}
\scalebox{0.67}{$\displaystyle
Z\!_{Q\!D\!V}\!=\!{C''}\!_{Q\!D\!V}\!\!\!\int\!\!{D{{\alpha}_i}}\exp\!\!\sum\limits_{\tau,\bm{x} }\!\left(\! \frac{-a_0}{\hbar }\! \right)\!\!{\left[\frac{1}{2\mu}{f^{ab}}{{\left( {\nabla'}\!\times\!\mathbf{\alpha} \right)}_a}\!{{\left( {\nabla'}\!\times\!\mathbf{\alpha} \right)}_b} \right]}\!\!\int\!\!D{\theta }\!\exp\!\!\left(\!\frac{-a_0}{\hbar }\!\right)
$}\nonumber\\
\scalebox{0.67}{$\displaystyle
\!\times\!\!\sum\limits_{\tau,\bm{x} }{\!\left\{ (E_J^0)_v\!\Bigl[ 1\!-\!\cos\!\left( {{\nabla }_0}{\theta }-2q{{\alpha}_0} \right) \Bigr]\!+\!(E_J)_v\!\sum\limits_{j=1}^2{\Bigl[ 1\!-\!\cos\!\left( {{\nabla }_j}{\theta }-2q{{\alpha}_j} \right) \Bigr]}\!\right\}}
$},
\end{align}
where \scalebox{0.9}{$\mu\!\equiv\!L/{\tilde{\cal{V}}_0\left( 0 \right)}$}, \scalebox{0.9}{$2q\equiv {2e}/{\hbar }\!=\!{2\pi}/{\Phi_0}$} represents a unit Cooper pair charge, which is twice the unit charge \scalebox{0.9}{$q\equiv{e}/{\hbar }\!=\!{2\pi}/{2{\Phi_0}}$}. Eq.(\ref{eq70}) shows that the pure $D\!A\!X\!Y$ model ($J\!J$ model without gauge coupling) has been dual transformed to the gauged $A\!X\!Y$ model (gauged $J\!J$ model) by the Villain approximation in the $2+1 d$ system. In other words, the gauge $J\!J$ model is a “frozen lattice dual superconductor” for the $Q\!P\!S$ model without gauge coupling.
\section{Mean field  analysis of  the gauged QPSJ model on the nanosheet \label{sec6}}
In this section, we introduce the mean field approximation to the partition function of Eq.(\ref{eq55}) and discuss its phase transition. From Eq. (\ref{eq55}), the partition function excluding the constant part is newly defined as ${Z_{G\!Q\!P\!J}}$of the gauged $Q\!P\!S\!J$ model, and using unit vectors of two real components ${\tilde{U}_i}=[\cos \tilde{\theta },\sin \tilde{\theta }]$, Eq. (\ref{eq55}) is rewritten as follows:
\begin{align}\label{eq72}
\scalebox{0.75}{$\displaystyle
Z_{G\!Q\!P\!J}\!=\!\!\int\!\!{D{{\tilde{\alpha }}_i}\left( x \right)}\exp \sum\limits_{x}{\left[ \frac{-1}{2{\tilde{\mu'}}}{f^{ab}}{{\left( {\nabla' }\times \mathbf{\tilde{\alpha }} \right)}_a}{{\left( {\nabla'}\!\times\!\mathbf{\tilde{\alpha}} \right)}_b}\left( x \right) \right]}
$}\nonumber\\ 
\scalebox{0.75}{$\displaystyle
\times \int{D\tilde{\theta }}\exp \sum\limits_{x}{\left\{{E'}\!_S\tilde{d}\sum\limits_{x}{\sum\limits_{l=1}^{2}{{\tilde{U}_l}\left( x \right){{\tilde{R}}_{\tilde{\alpha }}}{{\tilde{U}}_l}\left( x \right)}} \right\}}
$}
\end{align}
where ${\tilde{\mu' }}\equiv {\hbar {\tilde{\mu''}}}/{{a_0}}$, and the lattice difference operator ${{\tilde{R}}_{{\tilde{\alpha }}}}$ is defined as:
\begin{align}\label{eq73}
\scalebox{0.80}{$\displaystyle
{\tilde{R}}_{\tilde{\alpha }}\equiv 1+\frac{1}{2\tilde{d}}\left( \sum\limits_{i=1}^{2}{{\bar{\tilde{D}}_i}{\tilde{D}}_i}+\tilde{\gamma }{\bar{\tilde{D}}_{\tau }}{\tilde{D}}_{\tau } \right),\tilde{\gamma }\equiv\frac{{E'}\!_S^0}{{E'}\!_S}
 $},
\end{align}
where $\tilde{d}\equiv 2+\tilde{\gamma }$ represents anisotropic dimensional constants of the gauged $Q\!P\!S\!J$; and ${{\tilde{D}}_i}$ and ${\bar{\tilde{D}}}_i$ are forward and backward covariant lattice derivatives, respectively, and are defined, for example, for a complex field \scalebox{0.90}{$\tilde{U}={\tilde{U}}_1\!+\!i{\tilde{U}}_2$}, as follows:
\begin{align}\label{eq74}
\scalebox{0.80}{$\displaystyle
{{\tilde{D}}_i}\tilde{U}\left( \mathbf{x},\tau  \right)\equiv \tilde{U}\left( \mathbf{x}+\mathbf{i},\tau  \right){{e}^{-i{{q}_{m}}{{{\tilde{\alpha }}}_{i}}\left( \mathbf{x},\tau  \right)}}-\tilde{U}\left( \mathbf{x},\tau  \right), 
 $},\nonumber\\
\scalebox{0.80}{$\displaystyle                               
{\bar{\tilde{D}}_i}\tilde{U}\left( \mathbf{x},\tau  \right)\equiv \tilde{U}\left( \mathbf{x},\tau  \right)-\tilde{U}\left( \mathbf{x}-\mathbf{i},\tau  \right){{e}^{i{{q}_{m}}{{\tilde{\alpha }}_i}\left( \mathbf{x}-\mathbf{i},\tau  \right)}}
 $},
\end{align}
the same applies to ${\tilde{D}}_{\tau }$ and ${\bar{\tilde{D}}}_{\tau }$. In Eq.(\ref{eq72}), we introduce two sets of real two-component fields ${\tilde{u}}_l$ and ${\tilde{\psi }}_l$ ($l=1,2$) which satisfy the following identity:
\begin{align}\label{eq75}
\scalebox{0.80}{$\displaystyle
\int_{-\infty }^{\infty }{\text{d}{{\tilde{u}}_1}\text{d}{{\tilde{u}}_2}}\int_{-\infty }^{\infty }{\frac{\text{d}{{\tilde{\psi }}_1}\text{d}{{\tilde{\psi }}_2}}{{{\left( 2\pi i \right)}^2}}\exp \left\{ -{{\tilde{\psi }}_l}\left( {{\tilde{u}}_{1}}-{{\tilde{U}}_l} \right) \right\}=1}
 $},
\end{align}
\begin{align}\label{eq76}
\scalebox{0.8}{$\displaystyle
Z\!_{G\!Q\!P\!J}\!=\!\!\int\!\!\!{D{{\tilde{\alpha}}_i}}\!\exp\!\!\sum\limits_{x}\!\!{\left[ \frac{-1}{2{\tilde{\mu'}}}\!f^{ab}\!{{\left( {\nabla' }\!\times\! \mathbf{\tilde{\alpha}}\right)}_a}\!{{\left( {\nabla'}\!\times\!\mathbf{\tilde{\alpha}} \right)}_b}\!\right]}\!\prod\limits_{x}\!{\prod\limits_{l=1}^2{\int_{-\infty }^{\infty }\!\!\!{\text{d}{{\tilde{u}}_l}\!\frac{\text{d}{{\tilde{\psi }}_l}}{2\pi i}}}}
$}\nonumber\\
\scalebox{0.8}{$\displaystyle
\!\times\!\exp\!\sum\limits_{x}\!{\sum\limits_{l=1}^2\!{{\left[{E'}_S\tilde{d}{{\tilde{u}}_l}{{\tilde{R}}_{\tilde{\alpha }}}{{\tilde{u}}_l}-{{\tilde{\psi }}_l}{{\tilde{u}}_l}+\ln {I_0}( |{{\tilde{\psi }}_l}| ) \right]}}}
$},\quad
\end{align}
where we have defined the functional integrals of $\tilde{\theta }$ as follows:
\begin{align}\label{eq77}
\scalebox{0.80}{$\displaystyle
\prod\limits_{x}\!\!{\int_{-\pi }^{\pi }\!\!{\frac{d\tilde{\theta }}{2\pi }}}\!\exp\!\left\{ \sum\limits_{x}{\sum\limits_{l=1}^{2}{{{\tilde{\psi }}_l}{{\tilde{U}}_l}}} \right\}\!=\!\exp\!\sum\limits_{x,l}{\left\{ \ln {I_0}( |{{\tilde{\psi }}_l}| ) \right\}}
 $},
\end{align}
where ${I_0}(|\tilde{\psi }|)$ (\scalebox{0.90}{$| {\tilde{\psi }} |\equiv \sqrt{{{\tilde{\psi }}_1}^2+{{\tilde{\psi }}_2}^2}$} ) represents the modified Bessel functions of a zeroth-order integer. In Eq.(\ref{eq76}), performing the integrals over ${\tilde{u}}_l$ fields, we obtain the partition function by the complex field $\tilde{\psi }\equiv {{\tilde{\psi }}_1}+i{{\tilde{\psi }}_2}$ and ${{\tilde{\psi }}^{*}}\equiv {{\tilde{\psi }}_1}-i{{\tilde{\psi }}_2}$.
\begin{align}\label{eq78}
\scalebox{0.7}{$\displaystyle
 Z\!_{G\!Q\!P\!J}\!=\!\int{D{{\tilde{\alpha }}_i}}\prod\limits_{x}{\left( \int_{-\infty }^{\infty }{\frac{\text{d}\tilde{\psi }\text{d}{{\tilde{\psi }}^{*}}}{4\pi {{E'}_S}\tilde{d}}} \right)}\exp \left\{ -{F'}\left( \tilde{\psi },{{\tilde{\psi }}^{*}},{{\tilde{\alpha }}_i} \right) \right\}
$}\nonumber\\
\scalebox{0.7}{$\displaystyle
 {F'}\left( \tilde{\psi },{{\tilde{\psi }}^{*}},{{\tilde{\alpha }}_i} \right)\!\!\equiv\!\!\sum\limits_x{\left\{ \!\frac{1}{2{\tilde{\mu' }}}f^{ab}{{\left( {\nabla'}\!\times\! \mathbf{\tilde{\alpha }} \right)}_a}{{\left( {\nabla'}\!\times\!\mathbf{\tilde{\alpha }} \right)}_b}+\frac{1}{4{{E'}_S}d}{{| \hat{\tilde{\psi }}|}^2}\!\!\!-\!\ln {I_0}( | \hat{\tilde{\psi }} | ) \right\}}
$}\nonumber\\
\scalebox{0.9}{$\displaystyle
\hat{\tilde{\psi }}\left( x \right)\equiv {{\tilde{R}}_{{\tilde{\alpha }}}}^{\frac{1}{2}}\tilde{\psi }\left( x \right)
$},\quad\quad\quad\quad\quad
\end{align}
In Eq.(\ref{eq78}), since $\tilde{\psi }$ and ${\tilde{\psi }}^{*}$ can be regarded as the order parameters of the superinsulator (i.e., the disorder parameters of the superconductor), the non-dimensional free energy ${F'}(\tilde{\psi },{{\tilde{\psi }}^*},{\tilde{\alpha }}_i)$ can be Landau expanded for terms up to ${\left| \psi  \right|}^{4}$, ${\left| {D_i}\psi  \right|}^2$ and ${\left| {D_{\tau }}\psi  \right|}^2$ as follows\cite{ref25}:
\begin{align}\label{eq79}
\scalebox{0.75}{$\displaystyle
 {F'}_{DGL}(\tilde{\psi },{{\tilde{\psi }}^*},{{\tilde{\alpha }}_i})\!\equiv\!\sum\limits_x\!{\left\{ \frac{1}{2{\tilde{\mu'}}}{f^{ab}}{{\left( {\nabla }'\times \mathbf{\tilde{\alpha }} \right)}_a}{{\left( {\nabla' }\times \mathbf{\tilde{\alpha }} \right)}_b} \right.}
$}\quad\quad\quad\nonumber\\
\scalebox{0.75}{$\displaystyle
\left. 
\!+\!\frac{1}{8\tilde{d}}\left( \sum\limits_{i=1}^2{{| {D_i}\tilde{\psi } |}^2}\!+\!\tilde{\gamma }{{\left| {D_{\tau }}\tilde{\psi } \right|}^2} \right)\!+\!\frac{1}{4}\left( \frac{1}{{E'}_S\tilde{d}}\!-\!1 \right){{|\tilde{\psi }|}^2}\!+\!\frac{1}{64}{{| \tilde{\psi } |}^4} \right\}
$},
\end{align}
${F'}\!_{D\!G\!L}$is the non-dimensional DGL energy of the superinsulator or $Q\!P\!S\!J$ model on the nanosheet in $\tilde{d}\equiv 2+\tilde{\gamma }$ dimension at zero temperature. Therefore, the critical values ${E'}_S^{MF}$ according to the mean field approximation of $Q\!P\!S$ amplitude ${E'}_S$ are as follows:
\begin{align}\label{eq80}
\scalebox{0.76}{$\displaystyle
{E'}_{S}^{MF}\equiv \frac{1}{{\tilde{d}}}=\frac{1}{2+\tilde{\gamma }}
 $},
\end{align}
 The continuous limit in Eq.(\ref{eq79}), is as follows:\\
\scalebox{0.64}{$\displaystyle
F\!_{D\!G\!L}(\tilde{\psi },{{\tilde{\psi }}^*},{{\tilde{A}}_i})\!\equiv\!\!\int\!\!{d{x_0}}\!\!\int\!\!{{d^2}x}\!\left\{ \frac{f^{ab}}{2\tilde{\mu }}{{( {\partial' }\times \mathbf{\tilde{A}} )}_a}{{( {\partial' }\times \mathbf{\tilde{A}} )}_b}\!+\!\frac{1}{2m_{\Phi }}\sum\limits_{i=1}^2{{{\left| \left( -i\hbar{\partial }_i-{{\Phi }_0}{{\tilde{A}}_i} \right)\tilde{\psi } \right|}^2}}
\right.
$}\\
\scalebox{0.64}{$\displaystyle
 \left.\quad\quad\quad\quad\quad\quad\quad
 +\frac{\tilde{\gamma }}{2m_{\Phi }}{{\left| \left( -i\hbar{\partial }_0-{\Phi }_0{\tilde{A}}_0\right)\tilde{\psi } \right|}^2}+{\tilde{\alpha}}{\tilde{\varepsilon}}{{| {\tilde{\psi }} |}^2}+\tilde{\beta} {{| \tilde{\psi } |}^4} \right\}
$}
\begin{align}\label{eq81}
\scalebox{0.8}{$\displaystyle
\tilde{\alpha}\equiv \frac{\hbar }{4{a_0}{a^2}},\;\tilde{\varepsilon}\equiv \frac{{E'}_S^{M\!F}-{E'}_S}{{E'}_S},\; \tilde{\beta}\equiv \frac{\hbar }{64{a_0}{a^2}}
$},
\end{align}
where $\tilde{\mu }\!\!\equiv\!\!{C}/{a^2{{\cal{V}}_0}\left( 0 \right)}$, $m_{\Phi }\!\!\equiv\!\!4\tilde{d}\hbar {a_0}$is a pseudo-mass of magnetic flux having a dimension of $\left[ \text{J}\cdot {{\text{s}}^{2}} \right]$, and ${\tilde{A}}_{\mu }$ represents dual vector potentials having a dimension of $\left[\text{C}/\text{m}\right]$. Therefore, the order parameter $\tilde{\psi }$ of the superinsulator is gauge coupled to the U(1) dual gauge field ${\tilde{A}}_{\mu }$ by a unit magnetic charge $q_m\!\!\equiv \!\!2\pi/2e\!\!=\!\!{\Phi }_0/\hbar$, and when $\tilde{\psi }$ is in the condensed state, As shown in Figure 4, the pair of the Cooper pair and the anti-Cooper pair12, 13 is confined within the superinsulator by the electric flux-tubes. 
\begin{figure}[htbp]
 \begin{minipage}{1.0\hsize}
 \includegraphics[keepaspectratio, height=11mm]{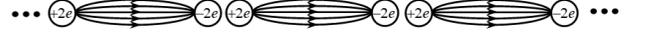}
\caption{Schematic diagram of a superinsulator on a nanosheet at zero temperature.}
\end{minipage}
\end{figure}
\\
Figure 4 shows the confinement of electric flux in the superinsulator between the Cooper pair and the anti-Cooper pair. Whether this confinement picture in which the charge of $2e$ for a superinsulator is the smallest unit of charge is correct will be discussed again in Section \ref{sec8} (Summary and discussion). From Eq.(\ref{eq81}), three DGL equations are derived as follows:
\begin{align}\label{eq82}
\scalebox{0.78}{$\displaystyle
{j_{\Phi }}^i\!=\!\frac{1}{\tilde{\mu }}{{\eta' }^{ia}}{{\left( {\partial' }\!\times\!\mathbf{\tilde{B}} \right)}_a}\!=\!\frac{\hbar{{\Phi }_0}}{2m_{\Phi }}\left\{ {{\tilde{\psi }}^*}{{\partial }_i}\tilde{\psi }\!-\!\left({\partial }_i{{\tilde{\psi }}^*} \right)\tilde{\psi } \right\}\!-\!\frac{{{\Phi }_0}^2}{m_{\Phi }}| \tilde{\psi } |^2\tilde{A}_i
 $},
\end{align}
\begin{align}\label{eq83}
\scalebox{0.76}{$\displaystyle
{j_{\Phi }}^0\!=\!\frac{1}{\tilde{\mu }}{{\eta' }^{0a}}{{\left( {\partial' }\!\times\!\mathbf{\tilde{B}} \right)}_a}\!=\!\frac{\gamma \hbar{{\Phi }_0}}{2m_{\Phi }}
\left\{ {{\tilde{\psi }}^*}{{\partial }_0}\tilde{\psi }-\left({\partial }_0{{\tilde{\psi }}^*}\right)\tilde{\psi }\right\}\!
-\!\frac{{{\Phi }_0}^2}{m_{\Phi }}| \tilde{\psi }|^2\tilde{A}_0
$},
\end{align}
\begin{align}\label{eq84}
\scalebox{0.76}{$\displaystyle
\frac{1}{2{m}_{\Phi }}\!\!\sum\limits_{i=1}^2\!{{{\left( \!-\!i\hbar {{\partial }_i}\!-\!{{\Phi }_0}{\tilde{A}_i}\right)}^2}\tilde{\psi }}\!+\!\frac{\gamma }{2{m}_{\Phi }}{{\left(\! -i\hbar {{\partial }_0}\!-\!{{\Phi }_0}{\tilde{A}_0} \right)}^2}\tilde{\psi }\!+\!2\beta |\tilde{\psi }|^2\tilde{\psi }\!=\!-\!\alpha \varepsilon \tilde{\psi }
$},
\end{align}
where ${\tilde{B}_i}\left( x \right)={\varepsilon_{iab}}{{\partial'}_a}{\tilde{A}_b}\left( x \right)$ represents a dual magnetic flux density. Eq.(\ref{eq82}) and (\ref{eq83}) represent the current density and Eq.(\ref{eq84}) represents the nonlinear Schrödinger equation for superinsulators. The thermodynamic critical dual magnetic field ${\tilde{H}_c}$ for the dual magnetic field ${\tilde{H}_0}={{{\tilde{B}}_0}}/{\tilde{\mu }}$ is as follows:
\begin{align}\label{eq85}
\scalebox{0.76}{$\displaystyle
\tilde{H}_c\left( {E'}_S\right)=\sqrt{\frac{{{\alpha }^2}{{\varepsilon }^2}}{2\beta \tilde{\mu }}}=\frac{\alpha }{\sqrt{2\beta \tilde{\mu }}}\left( 1-\frac{{E'}_S^{MF}}{{{E'}_S}} \right)
$},
\end{align}
where $\tilde{H}_c$ is the dimension of the voltage. Both have a dimension of length, the penetration depth $\tilde{\lambda }({E'}\!_S)$ and the coherent length $\tilde{\xi }({E'}\!_S)$ in the super insulator, which both have the dimension of length, are as follows:
\begin{align}\label{eq86}
\scalebox{0.85}{$\displaystyle
\tilde{\lambda }\left( {E'}\!_S \right)=\sqrt{\frac{\beta m_{\Phi }}{2\pi\tilde{\mu }\alpha{\Phi_0}^2 }}{{\left( \frac{{E'}\!_S}{{E'}_S-{E'}\!_S^{M\!F}} \right)}^{\frac{1}{2}}}
$},\quad\end{align}
\begin{align}\label{eq87}
\scalebox{0.85}{$\displaystyle
\tilde{\xi }\left( {E'}\!_{S}\right)=\frac{\hbar }{\sqrt{2{{m}_{\Phi }}\alpha }}{{\left( \frac{{E'}\!_S}{{E'}_S-{E'}\!_S^{M\!F}} \right)}^{\frac{1}{2}}}
$}.
\end{align}
From Eqs.(\ref{eq86}) and (\ref{eq87}), the dual Ginzburg–Landau parameter $\kappa$ is as follows:
\begin{align}\label{eq88}
\scalebox{0.80}{$\displaystyle
\tilde{\kappa }\left( {E'}_S \right)\!=\!\frac{\tilde{\lambda }\left( {E'}_S \right)}{\tilde{\xi }\left( {E'}_S \right)}=\frac{m_{\Phi }}{\hbar {{\Phi }_0}}\sqrt{\frac{\beta }{\pi \tilde{\mu }}}\!=\!\frac{2\pi }{2e}\sqrt{8\pi }\tilde{B}_c\!\left( {E'}_S \right){{\tilde{\lambda }}^2}\!\left( {E'}_S \right)
$},
\end{align}
where ${\tilde{B}_c}=\tilde{\mu }{\tilde{H}_c}$ represents the thermodynamic critical dual magnetic flux density. From the analogy of the classification of type I and type I\hspace{-.1em}I superconductors, the following classifications by type I and type I\hspace{-.1em}I superinsulators are formed from the sign of the surface energy ${\tilde{\sigma }_{SN}}$ and the value of the Ginzburg–Landau parameter $\kappa$ at the superinsulator–normal insulator boundary.\\
i)  type-I superinsulator    
\[ \tilde{\kappa }\textless\frac{1}{\sqrt{2}},\;\;\;\;\;\;{\tilde{\sigma }_{SN}}\textgreater0. \]
ii) intermediate of type I and type I\hspace{-.1em}I superinsulator 
\[ \tilde{\kappa }=\frac{1}{\sqrt{2}},\;\;\;\;\;\;{{\tilde{\sigma }}_{SN}}=0. \]
iii) typeI\hspace{-.1em}I superinsulator       
\[ \tilde{\kappa }\textgreater\frac{1}{\sqrt{2}},\;\;\;\;\;\;{{\tilde{\sigma }}_{SN}}\textless0. \]
From the analogy with the mixed state of the type-I\hspace{-.1em}I superconductor, the possibility of the existence of the mixed state in the case of the type-I\hspace{-.1em}I superinsulator is expected, and, from the analogy with the Abrikosov magnetic flux lattice of the type-I\hspace{-.1em}I  superconductor, the existence of an electric flux lattice is also expected in the case of the type-I\hspace{-.1em}I superinsulator on a nanosheet.
\section{Estimating the critical value of the QPS amplitude by the effective energy approach \label{sec7}}
In the previous section, we derived the thermodynamic critical dual magnetic field $\tilde{H}_c$, the penetration depth $\tilde{\lambda }$, and the coherent length $\tilde{\xi }$ from mean field analysis, all of which depended on the difference between the $Q\!P\!S$ amplitude \scalebox{0.9}{${E'}\!_S$} and its mean field critical value \scalebox{0.9}{${E'}\!_S^{M\!F}$}. Since the mean field approximation is a very rough approximation, in this section, we show the calculation result of the mean energy approximation with the contribution of fluctuations up to the two loop corrections by the effective energy approach shows in Appendix B.
The critical value $\left( {E'}\!_S \right)_c^{2\text{loop}}$ of the $Q\!P\!S$ amplitude ${E'}\!_S$ in the mean field approximation with up to the two loop corrections is given to Eq.(\ref{eqB14}) as a function of the anisotropy parameter $\tilde{\gamma }$. 
The results are plotted in Figure 5. Similarly, Figure 5 shows the results for the tricritical point ${{\left( {E'}\!_S \right)}^{tri}}$ of the $Q\!P\!S$ amplitude given in Eq.(\ref{eqB15}).
\begin{figure}[htbp]
 \begin{minipage}{1.0\hsize}
 \includegraphics[keepaspectratio, height=62mm]{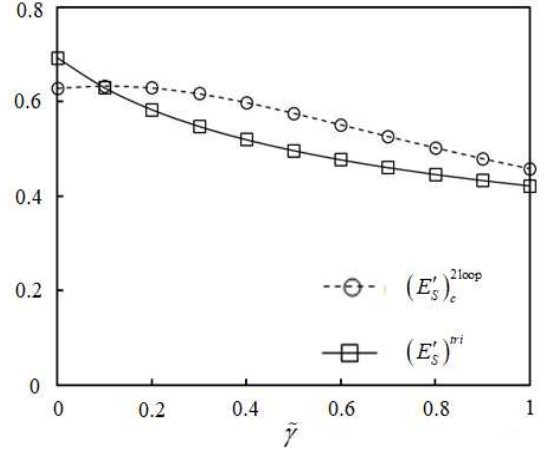}
\caption{The critical value and the tricritical point as a function of anisotropy parameter in mean field approximation with up to two loop corrections. Where, $\tilde{\gamma }$ is the anisotropy parameter, and ${{\left( {E'}\!_S \right)}_c^{2loop}}$ and ${{\left( {E'}\!_S \right)}^{tri}}$ are the critical value and the tricritical point of $Q\!P\!S$ amplitud, respectively.}
\end{minipage}
\end{figure}
From Eqs. (\ref{eq51}) and (\ref{eq11}), the $Q\!P\!S$ amplitude ${{E'}\!_S}$ has the following relationship with the charging energy ${{E'}\!_c}$ :
\begin{align}\label{eq89}
\scalebox{0.95}{$\displaystyle
{E'}\!_c=2{{\pi }^2}{{\cal{V}}_0}\left( 0 \right){{\left( {E'}\!_S \right)}_v}
$},
\end{align}
where ${{\cal{V}}_0}\left( 0 \right)$ is a massless lattice Green's function having an anisotropy parameter $\gamma$ in the $J\!J$ model defined by Eq.(\ref{eq43}).Using Eq.(\ref{eq69}), we have plotted in Figure 6 the critical value ${{\left( {E'}_c \right)}_c}$ of the charging energy ${E'}_c$ for various values of the anisotropy parameter $\gamma$ in $J\!J$ models as a function of $\tilde{\gamma }$

\begin{figure}[htbp]
 \begin{minipage}{1.0\hsize}
 \includegraphics[keepaspectratio, height=55mm]{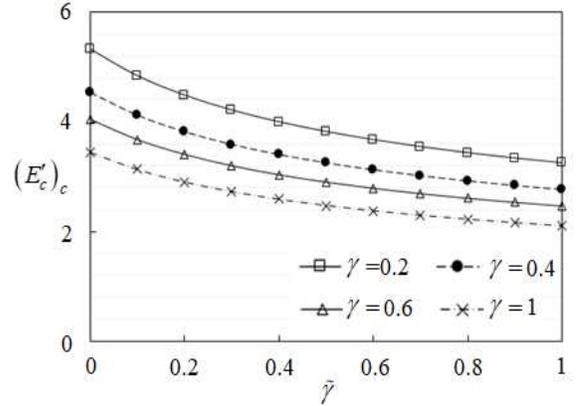}
\caption{The critical value of the charging energy for various values of the anisotropy parameter $\gamma$ in Josephson junction ($J\!J$) models as a function of $\tilde{\gamma }$.}
\end{minipage}
\end{figure}
\section{Summary and discussion \label{sec8}}
The conclusions of this paper are summarized below. First, using the dual Hamiltonian method, the phase and amplitude relationship between the $J\!J$ and $Q\!P\!S\!J$ models without gauge coupling on $2+1d$ nanosheets at zero temperature was determined, and the relationships between various constants were derived. Furthermore, the exact duality between the $J\!J$ model and the $Q\!P\!S\!J$ model based on the nonlinear Legendre transformation between the Lagrangian and the Hamiltonian using canonical conjugate variables of infinite order in a compact $2+1d$ lattice space was demonstrated. A dual transformation from the $A\!X\!Y$ model to the gauged $D\!A\!X\!Y$ model by the Villain approximation in the $2+1d$ system was derived.there are two main differences between the dual transformation by the dual Hamiltonian method and the dual transformation by the Villain approximation. One is that Eqs.(\ref{eq12}) and (\ref{eq51}) differ only in the core potential ${\cal{V}}_0\left( 0 \right)$. Another difference is that, in the case of the dual transformation by the Villain approximation, there is a gauge coupling by the dual gauge field ${{\tilde{\alpha' }}_i}$, however, in the dual transformation by the dual Hamiltonian method, there is no gauge coupling by the dual gauge field. The gauge coupling by the dual gauge field ${{\tilde{\alpha' }}_i}$ is gauge coupled with the $U(1)$ dual gauge field by the unit magnetic charge ${q_m}\!\equiv\!{2\pi}/{2e}$ by the scaling of $2e$ introduced in Eq.(\ref{eq54}), and, as shown in Figure 4, it was shown that the electric flux of the Cooper pair and the anti-Cooper pair in units of $2e$ was confined in the superinsulator. In the following, we consider whether the superinsulator's picture of confinement in $2e$ units that is, confinement by a pair consisting of a Cooper pair and an anti-Cooper pair is correct or incorrect. In the case of the magnetic flux confinement for the superconductor, the magnetic flux quantum ${{\Phi }_0}=h/{2e}$, which is the minimum unit of magnetic flux, is confined. However, if Eq.(55) is correct, in the confinement of charges in the superinsulator, $2e=h/{{{\Phi }_0}}$, which is twice the elementary charge $e=h/{\left( 2{{\Phi }_0} \right)}$,which is the minimum unit of charge, is confined. This is clearly inconsistent with the superconducting case. Therefore, in the case of a superinsulator, it should also be considered correct to assume that confinement occurs in units of the elementary charge $e\!=\!h/{\left( 2{{\Phi }_0} \right)}$, which is the minimum unit of charge. In other words, since the scaling by $2e$ in Eq.(\ref{eq54}) in Section 4 was completely artificial, one could change Eq. (54) to scaling by $e$ as follows:
\begin{align}\label{eq90}
\scalebox{0.95}{$\displaystyle
 e{{\tilde{\alpha'}}_i}\left( x \right)\equiv {{\tilde{\alpha }}_i}\left( x \right)
$},
\end{align}
By the transformation of Eq.(\ref{eq90}), Eq.(\ref{eq55}) is transformed as follows:
\begin{align}\label{eq91}
\scalebox{0.68}{$\displaystyle
Z\!_{QV}\!=\!{C''}\!_{QV}\!\!\!\int\!\!{D{\tilde{\alpha}\!_i}}\exp\!\!\sum\limits_{\tau,\bm{x}}\!\!\left(\! \frac{-a_0}{\hbar }\! \right)\!\!\!{\left[ \frac{1}{2{\tilde{\mu'' }}}{f^{ab}}{{\left( {\nabla}\times \mathbf{\tilde{\alpha}} \right)}_a}\!{{\left( {\nabla}\times \mathbf{\tilde{\alpha}} \right)}_b} \right]}\!\!\int\!\!D\tilde{\theta }\!\exp\!\left(\!\frac{-a_0}{\hbar }\!\right)
$}\nonumber\\
\scalebox{0.68}{$\displaystyle
\times\!\sum\limits_{\tau,\bm{x}}\!\left\{ E_S^0\!\left[ 1\!-\!\cos\!\left( {\nabla }_0\tilde{\theta }-2q_m {\tilde{\alpha}_0} \right) \right]\!+\!{E_S}\!\sum\limits_{j=1}^2{\left[ 1\!-\!\cos\!\left( {\nabla }_j\tilde{\theta }-2q_m{\tilde{\alpha}_j} \right) \right]}\!\right\}
$},
\end{align}
When Eq.(\ref{eq91}) is compared with Eq.(\ref{eq55}), the coupling magnetic charge is $2q_m\!=\!2{\Phi }_0/\hbar\!=\!2\pi/e$, which is twice the unit magnetic charge $q_m\!=\!{\Phi }_0/\hbar\!=\!2\pi/{2e}$. Figure 7 shows the confinement of the electric flux between the positive and the negative elementary charges in the superinsulator, that is, it differs from Figure 4.
\begin{figure}[htbp]
 \begin{minipage}{1.0\hsize}
 \includegraphics[keepaspectratio, height=12mm]{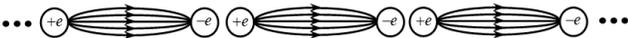}
\caption{Schematic diagram of a superinsulator with an elementary charge $e$ on a nanosheet at zero temperature.}
\end{minipage}
\end{figure}
To obtain the result of Eq.(\ref{eq91}), $E_c$ introduced in Eq.(\ref{eq1}) is not ${\left( 2e \right)}^2/2C$ but rather $e^2/{2C}$. In this case, the $D\!G\!L$ free energy for Eq.(\ref{eq91}) is as follows:
\scalebox{0.67}{$\displaystyle
F\!_{D\!G\!L}(\tilde{\psi },{{\tilde{\psi }}^*},{{\tilde{A}}_i})\!\!\equiv\!\!\!\int\!\!{d{x_0}}\!\!\int\!\!{{d^2}\!x}\!\left\{ \frac{f^{ab}}{2\tilde{\mu }}{{( {\partial' }\!\times\! \mathbf{\tilde{A}} )}_a}{{( {\partial' }\!\times\!\mathbf{\tilde{A}} )}_b}\!+\!\frac{1}{2m_{\Phi }}\!\sum\limits_{i=1}^2\!{{{\left| \left( \!-i\hbar{\partial }_i\!-\!2{{\Phi }_0}{{\tilde{A}}_i} \right)\tilde{\psi } \right|}^2}}
\right.
$}
\begin{align}\label{eq92}
\scalebox{0.67}{$\displaystyle
 \left.
 +\frac{\tilde{\gamma }}{2m_{\Phi }}{{\left| \left( \!-i\hbar{\partial }_0\!-\!2{\Phi }_0{\tilde{A}}_0\right)\tilde{\psi } \right|}^2}\!+\!{\tilde{\alpha}}\tilde{\varepsilon}{{| {\tilde{\psi }} |}^2}+{\tilde{\beta}}{{| \tilde{\psi } |}^4} \right\}
$}
\end{align}
where ${{m}_{2\Phi }}$ is the effective mass of the flux pair (vortex pair in the same rotation direction = superinsulator) and $\tilde{\psi }$ is the wave function of the flux pair. On the other hand, the Ginzburg–Landau free energy for Eq.(\ref{eq71}) is as follows:
\scalebox{0.64}{$\displaystyle
F\!_{G\!L}(\psi ,{\psi }^*,A_i)\!\equiv\!\!\int\!\!{d{x_0}}\!\!\int\!\!{{d^2}x}\!\left\{ \frac{{\tilde{f}}^{ab}}{2\mu}{{( {\partial' }\times \mathbf{A} )}_a}{{( {\partial' }\times \mathbf{A} )}_b}\!+\!\frac{1}{2m_c}\sum\limits_{i=1}^2{{{\left| \left( -i\hbar{\partial }_i-{{\Phi }_0}{A_i} \right)\tilde{\psi } \right|}^2}}
\right.
$}\\
\scalebox{0.64}{$\displaystyle
 \left.\quad\quad\quad\quad\quad\quad\quad
 +\frac{\gamma}{2m_c}{{\left| \left( -i\hbar{\partial }_0-{\Phi }_0{A}_0\right)\tilde{\psi } \right|}^2}+\tilde{\alpha}\varepsilon{{| {\psi } |}^2}+\tilde{\beta}{{| \psi  |}^4} \right\}
$}
\begin{align}\label{eq93}
\scalebox{0.8}{$\displaystyle
\varepsilon \equiv \frac{{E'}_J^{M\!F}-{E'}_J}{{E'}_J}
$},
\end{align}
where $m_c$  is the effective mass of the Cooper pair and  $\psi$ is the wave function of the Cooper pair. It is known that the microscopic theory of superconductivity with respect to the Ginzburg–Landau theory can be described by the following BCS Hamiltonian ($H_{B\!C\!S}$):
\begin{align}\label{eq94}
\scalebox{0.8}{$\displaystyle
H\!_{B\!C\!S}\!\equiv\!\!\!\sum\limits_{\sigma =\uparrow ,\downarrow }\!{\int\limits_{\Omega }{\!{d^3}x\varphi _{\sigma }^{+}\!\left( x \right)}}\!\left\{ \frac{1}{2{m_e}}{\left[ i\hbar {{\partial }_j}\!-\!e{A_j}\!\left( x \right) \right]^2}\!-\!\mu\right\}\!\varphi_{\sigma }\!\left( x \right)
$}\nonumber\\
\scalebox{0.8}{$\displaystyle
-\frac{\left| g \right|}{2}\!\sum\limits_{\sigma ,{\sigma'}}\!{\int\limits_{\Omega }{\!{d^3}x}}\varphi _{\sigma }^{+}\!\left( x \right)\varphi_{\sigma'}^{+}\!\left( x \right)\varphi _{\sigma'}\!\left( x \right)\varphi_{\sigma }\!\left( x \right)
$},\quad
\end{align}
where ${m_e}$ is the effective mass of the electron, and ${{\varphi }_{\sigma }}\left( x \right)$ is the electron field having a spin subscript $\sigma$ and is a fermion satisfying the anti-commutation relation
\[{{\left\{ {{\varphi }_{\sigma }}\left( x \right),\varphi _{{\sigma'}}^+\left( x' \right) \right\}}_+}={{\delta }_{\sigma {\sigma'}}}{{\delta }^3}\left( x-x' \right).\]
From Eqs.(\ref{eq93}) and (\ref{eq94}), in microscopic theory, the matter field ${\varphi }_{\sigma }$is a fermion field which is coupled to the gauge field by an elementary charge $e$. On the other hand, in the Ginzburg–Landau theory, the material field $\tilde{\psi }$ is a boson field which is coupled to the gauge field by $2e$. The magnetic flux quantum in the superconductor have a repulsive force, form a vortex lattice, and never intersect with each other, so they can be regarded as elementary excitations of fermions.Therefore, from the analogy between the BCS theory and the Ginzburg–Landau theory of superconductivity, it is expected that the microscopic theory of superinsulators with respect to the DGL theory in Eq.(\ref{eq92}) can be described by the following dual BCS Hamiltonian:
\begin{align}\label{eq95}
\scalebox{0.8}{$\displaystyle
H\!_{D\!B\!C\!S}\!\equiv\!\!\!\sum\limits_{\sigma =\uparrow ,\downarrow }\!{\int\limits_{\Omega }{\!{d^3}x\tilde{\varphi} _{\tilde{\sigma} }^{+}\!\left( x \right)}}\!\left\{ \frac{1}{2{m_{\Phi }}}{\left[ i\hbar {{\partial }_j}\!-\!{\Phi }_0{\tilde{A}_j}\!\left( x \right) \right]^2}\!-\!\tilde{\mu}\right\}\!\tilde{\varphi}_{\tilde{\sigma} }\!\left( x \right)
$}\nonumber\\
\scalebox{0.8}{$\displaystyle
-\frac{\left| \tilde{g} \right|}{2}\!\sum\limits_{\tilde{\sigma} ,{\tilde{\sigma}'}}\!{\int\limits_{\Omega }{\!{d^3}x}}\tilde{\varphi}_{\tilde{\sigma} }^+\!\left( x \right)\tilde{\varphi}_{\tilde{\sigma}'}^+\!\left( x \right)\tilde{\varphi}_{\tilde{\sigma}'}\!\left( x \right)\tilde{\varphi}_{\tilde{\sigma} }\!\left( x \right)
$},\quad
\end{align}
where ${m_{\Phi }}$ is the magnetic flux (vortex) quantum, and ${{\tilde{\varphi }}_{\sigma }}$ is the magnetic flux quantum field having the pseudo-spin subscript $\tilde{\sigma}$, and is a fermion satisfying the anti-commutation relationship
\[{{\left\{ {{\tilde{\varphi }}_{\tilde{\sigma} }}\left( x \right),\tilde{\varphi }_{{\tilde{\sigma}'}}^{+}\left( x' \right) \right\}}_+}={{\delta }_{\tilde{\sigma} {\tilde{\sigma}' }}}{{\delta }^3}\left( x-x' \right).\]
Similar to the relationship between Eqs.(\ref{eq93}) and (\ref{eq94}), in the microscopic theory of superinsulators of Eq.(\ref{eq94}), the material field ${{\tilde{\varphi }}_{\sigma }}$ is a fermion field which is coupled to the gauge field by the flux quantum ${{\Phi }_0}$. On the other hand, in the DGL theory of Eq.(\ref{eq92}), the material field $\tilde{\psi }$ is a boson field and is coupled to the gauge field by 2${{\Phi }_0}$. The construction of the dual BCS theory, which is a microscopic theory of superinsulators on nanosheets, not only elucidates the microscopic mechanisms for superinsulators and quantum phase slips, but also the microscopic mechanisms for quark confinement and asymptotic freedom. This theory is expected to play an important role as a powerful model of the QCD phenomenon on desktop\cite{ref12,ref13}.
\section{Acknowledgments\label{Acknowledgments}}
I would like to thank all the faculty and staff of Aichi University of Technology.
\appendix

\section{Anisotropic lattice Green's function ${\cal{V}}_m\left( \mathbf{0} \right)$ at the source $x=0$}\label{appa}
Perform a numerical evaluation of the anisotropic massive lattice Green's function (lattice potential)${{\cal{V}}_m}\left( \mathbf{0} \right)$ at the origin $x=0$.
\begin{align}\label{eqA1}
\scalebox{1.0}{$
{\cal{V}}_m\left( \mathbf{0} \right)\!\equiv\!\frac{1}{-g^{\mu \nu }{\bar{\nabla }}_{\mu }{\nabla }_{\nu }+m^2}=\sum\limits_{n=0}^{\infty }\frac{h_n}{{\left[ \;m^2\;+\;2\tilde{d}\; \right]}^{n+1}}
$},
\end{align}
where, $\tilde{d}\equiv2+\tilde{\gamma }$ is the anisotropic dimension, and $h_n$ is the anisotropic hopping coefficient of the anisotropic massive lattice Green's function ${\cal{V}}_m\left( \mathbf{0} \right)$ at the origin $x=0$, and is introduced as follows:
\begin{align}\label{eqA2}
\scalebox{1.0}{$
 h_n\equiv n!\sum\limits_{j=0,2,4}^n{\frac{{{\tilde{\gamma }}^{n-j}}}{{{\left\{\left[ \frac{\left( n-j \right)}{2}\right]! \right\}}^2}j!}}{H_j}
$},
\end{align}
where $H_n$ represents the isotropic hopping coefficients\cite{ref25}, for example, in the two-dimensional case, $H_0=1$, $H_2=4$, $H_6=36$, $H_8=400$,…, and in the three-dimensional case, $H_0=1$, $H_2=6$, $H_6=90$, $H_8=1860$,…,. TABLE$\rm{\,I\,}$ lists examples of $\tilde{\gamma }=0.1, 0.2, ...., 0.8, 0.9$, and $1.0$. 
\begin{table}[htbp]
\caption{\label{tab:table1}%
Values of the anisotropic hopping coefficient $h_n$ for values of n up to 10 for values of the anisotropic parameter $\tilde{\gamma }=0$ to 1 in the case of $2< \tilde{d} \leq 3$, where $\tilde{d}\equiv2+\tilde{\gamma }$.}
\begin{tabular}{cccccc} \hline\hline
$\tilde{\gamma }$ &  $h_2$ &  $h_4$  &  $h_6$  & $h_8$ & $h_{10}$ \\ \hline
\;\;0\;\;  & 4    & 36     & 400        & 4900        & 63504       \\
\;\;0.1\;\;  & 4.02 & 36.003 & 410.83602  & 5125.514241 & 67964.55133 \\
\;\;0.2\;\;  & 4.08 & 36.048 & 443.77728  & 5820.335539 & 81960.10908 \\
\;\;0.3\;\; & 4.18 & 36.243 & 500.13058  & 7040.109553 & 107387.453  \\
\;\;0.4\;\;  & 4.32 & 36.768 & 582.09792  & 8880.292915 & 147593.9992 \\
\;\;0.5\;\;  & 4.5  & 37.875 & 692.8125   & 11480.27344 & 207665.9648 \\
\;\;0.6\;\;  & 4.72 & 39.888 & 836.38912  & 15029.23717 & 294849.9426 \\
\;\;0.7\;\;  & 4.98 & 43.203 & 1017.98898 & 19773.88112 & 419126.4121 \\
\;\;0.8\;\; & 5.28 & 48.288 & 1243.89888 & 26028.09861 & 593959.5603 \\
\;\;0.9\;\;  & 5.62 & 55.683 & 1521.62482 & 34184.79254 & 837254.033  \\
\;\;1.0\;\;  & 6    & 66     & 1860       & 44730       & 1172556 \\ \hline\hline 
\end{tabular}
\end{table}
The asymptotic behavior of the anisotropic massive lattice Green's function ${\cal{V}}_m\!\left( \mathbf{0} \right)$ at small values of $m$ in the case of $2< d \leq 3$ is as follows:
\begin{align}\label{eqA3}
\scalebox{0.7}{$\displaystyle
{\cal{V}}_m\left( \mathbf{0} \right)=\frac{{{\tilde{\gamma }}^{-1/2}}\sqrt{2\tilde{d}}}{4\pi }\frac{\sqrt{{m^2}\!+\!4\tilde{d}}\!-\!\sqrt{m^2}}{\sqrt{m^2+2\tilde{d}}}\!+\!\!\!\sum\limits_{n=0,2,4..}^{\infty }{\frac{\Delta {h_n}}{{\left( m^2+2\tilde{d} \right)}^{n+1}}}
$}\nonumber\\
\scalebox{0.7}{$\displaystyle
\Delta {h_n}\!\equiv\!n!\!\!\!\sum\limits_{j=0,2,4}^n\!\!{\frac{{{\tilde{\gamma }}^{n-j}}}{{{\left\{ \left[ {\left( n-j \right)}/{2}\; \right]! \right\}}^2}j!}{H_j}}\!-\!\frac{{{\tilde{\gamma }}^{-1/2}}\sqrt{2\tilde{d}}}{4\pi }\frac{\left( 2n \right)!}{{2^{2n}}{{\left( n! \right)}^2}}\frac{{{\left( 2\tilde{d} \right)}^{n+1}}}{n+1}
$},
\end{align}
TABLE$\rm{\,I\hspace{-.01em}I\,}$ shows the asymptotic anisotropic hopping coefficient $\Delta {h_n}$ for values of $n$ up to 10 for values of the anisotropic parameter $\tilde{\gamma }=0$ to 1 in the case of $2< d \leq 3$.
\begin{table}[htbp]
\caption{\label{tab:table2}
Asymptotic anisotropic hopping coefficient $\Delta{h_n}$ for values of n up to 10 for values of the anisotropic parameter $\tilde{\gamma }=0$ to 1 in the case of $2< \tilde{d} \leq 3$, where $\tilde{d}\equiv2+\tilde{\gamma }$.
}
\begin{tabular}{cccccc} \hline\hline
$\tilde{\gamma }$ & $\Delta{h_2}$ & $\Delta{h_4}$ & $\Delta{h_6}$ & $\Delta{h_8}$ & $\Delta{h_{10}}$\\ \hline
\;\;0.1\;\; &-1.16603 & -0.7561 & -0.3789  & 27.6816  &549.214\\
\;\;0.2\;\; & -0.64231 & 0.10562 & 4.26659 & 59.7304 & 786.122\\
\;\;0.3\;\; & -0.4334 & 0.38867 & 5.27033 & 62.4803 & 769.857 \\
\;\;0.4\;\; & -0.32319 &  0.50921 &  5.42088 & 60.563 & 744.348\\
\;\;0.5\;\; & -0.25823 & 0.56803 & 5.36908 & 59.2432  & 755.792\\
\;\;0.6\;\; & -0.2182 & 0.60248  & 5.34728 & 60.225  & 818.965 \\
\;\;0.7\;\; & -0.19353 & 0.6296 & 5.46036 & 64.2966 &  944.418\\
\;\;0.8\;\; & -0.17904 & 0.65818 & 5.7662 & 72.0562 & 1145.97\\
\;\;0.9\;\; & -0.17169 & 0.69306 & 6.30443 & 84.1746 & 1443.87\\
\;\;1.0\;\; & -0.16955 & 0.73705 & 7.10848 & 101.516 & 1866.95\\ \hline\hline  
\end{tabular}
\end{table}
From Eq.(\ref{eqA3}) and TABLE$\rm{\,I\hspace{-.01em}I\,}$, when the value of the anisotropic massless lattice Green's function ${\cal{V}}_0\left( \mathbf{0} \right)$ at the origin $x\!=\!0$ is evaluated as the sum of the power-series up to $n\!=\!10$, it becomes as shown in TABLE$\rm{\,I\hspace{-.15em}I\hspace{-.15em}I\,}$ and Figure 6: 
\begin{table}[htbp]
\caption{\label{tab:table3}
Asymptotic massless lattice Green's function ${\cal{V}}_0\left( \mathbf{0} \right)$ at the origin $x\!=\!0$ for values of $n$ up to 10 for values of the anisotropic parameter $\tilde{\gamma }=0$ to 1 in the case of $2< d \leq 3$, where $\tilde{d}\equiv2+\tilde{\gamma }$.
}
\begin{tabular}{cc} \hline\hline
$\tilde{\gamma }$ & ${\cal{V}}_0\left( \mathbf{0} \right)$\\ \hline
\;\;\;0.1\;\;\; & \;0.444983243\; \\
\;\;\;0.2\;\;\; & \;0.389718363\; \\
\;\;\;0.3\;\;\; & \;0.355795798\; \\
\;\;\;0.4\;\;\; & \;0.331151594\; \\
\;\;\;0.5\;\;\; & \;0.311864992\; \\
\;\;\;0.6\;\;\; & \;0.29608701\; \\
\;\;\;0.7\;\;\; & \;0.282786758\; \\
\;\;\;0.8\;\;\; & \;0.271328982\; \\
\;\;\;0.9\;\;\; & \;0.261294231\; \\
\;\;\;1.0\;\;\; & \;0.252390927\; \\ \hline\hline  
\end{tabular}
\end{table}
\begin{figure}[htbp]
 \begin{minipage}{1.0\hsize}
 \includegraphics[keepaspectratio, height=55mm]{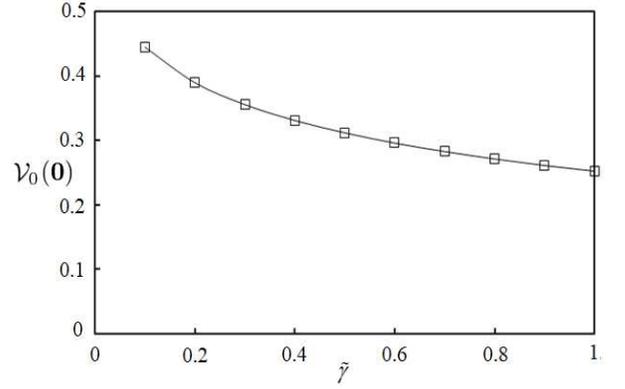}
\caption{Asymptotic massless lattice Green's function ${\cal{V}}_0\left(\mathbf{0}\right)$ at the origin $x=0$ versus the anisotropic parameter $\tilde{\gamma }$.}
\end{minipage}
\end{figure}
\section{ Appendix B.   Effective energy approach for the $Q\!P\!S\!J$ model}\label{appb}
Consider the order parameter representation of the super insulator by the partition function of Eq.(\ref{eq78}). To simplify the problem, we deal with the case where there is no coupling of gauge fields. First, to derive the one-loop effective field theory, the free correlation function in the presence of a non-vanishing background field is shown below:
\begin{align}\label{eqB1}
\scalebox{0.85}{$\displaystyle
{{\left. G_{{\tilde{\psi }}}^{-1}{{\left(x_1,x_2\right)}_{ab}} \right|}_{\hat{\tilde{\psi }}\!=\!\alpha }}\!=\!{{\left. V_2^{{\tilde{\psi }}}{{\left(x_1,x_2 \right)}_{ab}} \right|}_{\hat{\tilde{\psi }}=\alpha }}
$}\quad\quad\quad\quad\nonumber\\
\scalebox{0.9}{$\displaystyle
\!=\!{{\left. {{E'}_S}\frac{{{\delta }^2}{F'}}{\delta {{\hat{\tilde{\psi }}}_a}\left( x_1 \right)\delta {{\hat{\tilde{\psi }}}_b}\left( x_2 \right)} \right|}_{\hat{\tilde{\psi }}\!=\!\alpha }}, \;\left\langle{{{\hat{\tilde{\psi }}}}_l}\left( x \right) \right\rangle \equiv {{\tilde{\alpha }}_l}
$},
\end{align}
where ${\alpha }_l$ is an expectation of \scalebox{0.9}{${\hat{\tilde{\psi }}}_l\left( x \right)$}, and, in general, all $n$-th order one-particle irreducible graphs involving the vertex functions can be computed as follows:
\begin{align}\label{eqB2}
\scalebox{0.75}{$\displaystyle
{{\left. V_n^{{\tilde{\psi }}}{{\left( x_1,x_2\cdot \cdot \cdot,x_n \right)}_{ab}} \right|}_{a_1,a_2\cdot \cdot \cdot ,a_n}}\!\!\!\!\!\!\!\!\!\!\!=\!\!{{\left. {{E'}\!_S}\frac{{{\delta }^n}{F'}}{\delta {{{\hat{\tilde{\psi }}}}_{a_1}}\!\left( x_1 \right)\delta {{{\hat{\tilde{\psi }}}}_{a_2}}\!\left( x_2 \right)\cdot \cdot \cdot \delta {{{\hat{\tilde{\psi }}}}_{a_n}}\!\left( x_n \right)} \right|}_{\psi =\alpha }}\!\!\!\!\!\!\!\!\!
$},
\end{align}
The \scalebox{0.9}{$2\!\times\!2$} matrix \scalebox{0.9}{$G_{{\tilde{\psi }}}^{-1}{\!\left( x_1,x_2 \right)}_{ab}$} defined in Eq.(\ref{eqA1}) is divided into longitudinal \scalebox{0.9}{$G_{\psi }^{-1}{\!\left( x,y \right)}^L$} and transverse \scalebox{0.9}{$G_{\psi }^{-1}{\!\left( x,y \right)}^T$} parts, which are respectively parallel and orthogonal to the expected value of the field \scalebox{0.9}{${\alpha}_l$}, as follows: 
\begin{align}\label{eqB3}
\scalebox{0.78}{$\displaystyle
G_{\tilde{\psi }}^{-1}{\!\left(x_1,x_2 \right)}_{ab}\!=\!\frac{1}{2{E'}\!_S\tilde{d}}{{\delta }_{ab}}{{\delta }_{x_1,x_2}}\!\!-\!\!\left( 1+\frac{1}{2\tilde{d}}{{g}^{\mu \nu }}{{\bar{\nabla }}_{\mu }}{{\nabla }_{\nu }} \right)\!\!\left(x_1,x_2 \right){{\eta }_{ab}}
$}\nonumber\\
\scalebox{0.78}{$\displaystyle
=G_{{\tilde{\psi }}}^{-1}{{\left( x_1,x_2 \right)}^L}P_{ab}^L+G_{{\tilde{\psi }}}^{-1}{{\left( x_1,x_2 \right)}^T}P_{ab}^T
$},\quad\quad\quad\nonumber\\
\scalebox{0.78}{$\displaystyle
G_{\psi }^{-1}{{\!\left( x,y \right)}^L}\!\equiv\!\frac{1}{2{E'}\!_S\tilde{d}}\!\left\{ \!{\delta }_{xy}\!-\!2{E'}\!_S\tilde{d}{{\eta }_L}\!\left( \!1\!+\!\frac{1}{2\tilde{d}}{g^{\mu \nu }}{{{\bar{\nabla }}}_{\mu }}{{\nabla }_{\nu }} \right)\!\!\left( x,y \right)\! \right\}
$},\nonumber\\
\scalebox{0.78}{$\displaystyle
G_{\psi }^{-1}{{\!\left( x,y \right)}^T}\!\equiv\!\frac{1}{2{E'}\!_S\tilde{d}}\!\left\{ \!{\delta }_{xy}\!-\!2{E'}\!_S\tilde{d}{{\eta }_T}\!\left( \!1\!+\!\frac{1}{2\tilde{d}}{g^{\mu \nu }}{{{\bar{\nabla }}}_{\mu }}{{\nabla }_{\nu }} \right)\!\!\left( x,y \right)\! \right\}
$},
\end{align}
where \scalebox{0.9}{${\eta }_{ab}$} is a \scalebox{0.9}{$2\!\times\!2$} matrix; and \scalebox{0.9}{$P_{ab}^L$} and \scalebox{0.9}{$P_{ab}^T$} are longitudinal and transverse projection matrices, respectively: 
\begin{align}\label{eqB4}
\scalebox{0.78}{$\displaystyle
{\eta }_{ab}=P_{ab}^L{\eta }_L+P_{ab}^T{\eta }_T,\;P_{ab}^T\equiv {\delta }_{ab}-\frac{{\tilde{\alpha}}_a{\tilde{\alpha }}_b}{{\left| {\tilde{\alpha }} \right|}^2},\;P_{ab}^L\equiv\frac{{\tilde{\alpha}}_a{\tilde{\alpha}}_b}{{\left| \tilde{\alpha } \right|}^2},
$}\nonumber\\
\scalebox{0.78}{$\displaystyle
Q\left( \left| {\tilde{\alpha }} \right| \right)\equiv\log {I_0}\left( \left| {\tilde{\alpha }} \right| \right),\; {{\eta }_{ab}}\equiv \frac{{{\delta }^2}Q\left( \left| {\tilde{\alpha }} \right| \right)}{\delta {{\tilde{\alpha }}_a}\delta {{\tilde{\alpha }}_b}},
$}\nonumber\\
\scalebox{0.78}{$\displaystyle
{{\eta }_{T}}\equiv\frac{1}{\left| \tilde{\alpha }  \right|}\frac{dQ\left( {{\left| \tilde{\alpha }  \right|}} \right)}{d\left| \tilde{\alpha } \right|},\;{{\eta }_{L}}\equiv\frac{{{d}^{2}}Q}{d{{\left| \tilde{\alpha } \right|}^{2}}},
$}
\end{align}
By integrating over all quadratic fluctuations in the partition function of Eq.(\ref{eq78}), the one-loop effective energy  is as follows:
\begin{align}\label{eqB5}
\scalebox{0.78}{$\displaystyle
{F'}^{1loop}=\frac{1}{2M{M_{\tau }}}\left\{Tr\log \left( \frac{-1}{2\tilde{d}}{g^{\mu \nu }}{\bar{\nabla }}_{\mu }{\nabla }_{\nu } \right) 
\right.
$}\nonumber\\
\scalebox{0.78}{$\displaystyle
\left. 
+Tr\log \left[ \frac{m^2}{2\tilde{d}}\!-\!\left( 1-\frac{m^2}{2\tilde{d}} \right)\frac{1}{2\tilde{d}}{g^{\mu \nu }}{{\bar{\nabla }}_{\mu }}{{\nabla }_{\nu }} \right]  
\right\}
 $},\nonumber\\
\scalebox{0.78}{$\displaystyle
m^2\equiv 4\tilde{d}\!-\!{\left( 2\tilde{d} \right)}^2\!{E'}\!_S\!\left\{ 1\!-\!\left[ \frac{I_1\left( {E'}_S \right)}{I_0\left( {E'}_S \right)}\right]^2\right\}
 $}, 
\end{align}
where $I_0\left( E \right)$ and $I_1\left( E \right)$ are order zero and order one the modified Bessel functions, respectively. The first trace means zero mass fluctuations (Goldstone modes) and the second trace represents massive fluctuations and can be calculated by hopping expansion as follows:
\begin{align}\label{eqB6}
\scalebox{0.78}{$\displaystyle
Tr\log \left[ \frac{m^2}{2\tilde{d}}\!-\!\left( 1-\frac{m^2}{2\tilde{d}} \right)\frac{1}{2\tilde{d}}{g^{\mu \nu }}{{\bar{\nabla }}_{\mu }}{{\nabla }_{\nu }} \right]  
 $}\nonumber\\
\scalebox{0.78}{$\displaystyle
=-\sum\limits_{n=2}^{\infty }{\frac{1}{n}}{{\left( 1-\frac{m^2}{2\tilde{d}} \right)}^n}\frac{h_n}{( 2\tilde{d} )^n}
 $}, 
\end{align}
where $h_n$ is the anisotropic hopping coefficient of the anisotropic massive lattice Green's function $V_m\left( \mathbf{0} \right)$ at the source $x=0$ shown in Appendix A. The one-loop free energy can be expressed by the hopping expansion of Eq.(\ref{eqB6}) as follows:
\begin{align}\label{eqB7}
\scalebox{0.78}{$\displaystyle
{F'}^{1loop}\!=\!\frac{-1}{2}\!\!\!\sum\limits_{n=0,2,4,...}^{\infty }\!\!{\frac{h_n}{n}}\left[ {{\left( \frac{\tilde{b}{\eta }_L}{\tilde{d}} \right)}^n}+{{\left( \frac{\tilde{b}{\eta }_T}{\tilde{d}} \right)}^n} \right]  
 $}, 
\end{align}
where $\tilde{b}\equiv{E'}\!_S\tilde{d}$. Next, from the one-particle irreducible diagrams $\bigcirc\!\bigcirc$ and \scalebox{1.4}{$\ominus$}, the free energy of the two loop corrections is derived as follows, respectively:
\begin{align}\label{eqB8}
\scalebox{0.78}{$\displaystyle
{F'}^{\bigcirc\!\bigcirc }\!=\!\frac{-1}{4!}\left\{4\ddot{Q}\!\left(\left| \tilde{\alpha } \right|\right)\!\!\left(3{\Bigl[\mathrm{Tr}G\!\left(0\right)\Bigr]}^2\!+\!6{{\Bigl[\mathrm{Tr}G\!\left(0\right)\Bigr]}^2} \right)\!+\!8\dddot{Q}\left(\left| \tilde{\alpha }  \right| \right)
\right.
 $}\nonumber\\
\scalebox{0.78}{$\displaystyle
\left. 
\times\left( 12{\bar{\hat{G}}}^2\!\!\left( 0 \right)+6\left[ \mathrm{Tr}\hat{G}\left( 0 \right) \right]\bar{\hat{G}}\left( 0 \right) \right)+16\ddddot{Q}\!\left( |\tilde{\alpha } | \right)3{{\left[ \bar{\hat{G}}\left( 0 \right) \right]}^2}
\right\}   
 $}, 
\end{align}
\begin{align}\label{eqB9}
\scalebox{0.78}{$\displaystyle
{F'}^{\scalebox{1.2}{$\ominus$} }\!\!=\!\frac{1}{2}{{\left( \!\frac{1}{3!} \right)}^2}\!\!\sum\limits_x\!\!\left\{\!{\left( 4\ddot{Q} \right)}^2\!\!\!+\!8\dddot{Q}\!\left(\left| \tilde{\alpha }\right| \right)\!\left[ 18\bar{\hat{G}}\!\left( x \right)\!\mathrm{Tr}\!\left[ \hat{G}{\left( x \right)}^2 \right]\!\!+\!36\bar{\hat{G}}\!\left( x \right) \right]
\right.
 $}\nonumber\\
\scalebox{0.78}{$\displaystyle
\left. 
\!+36\!\left(4\ddot{Q}\right)\!\!\Bigl(8\dddot{Q}\Bigr)\!\!\left[ {\bar{\hat{G}}}^2\!\!\!\left( x \right) \right]\!\bar{\hat{G}}\!\left( x \right)\!+\!6\left(8\dddot{Q}\right)^2\!{{\left[ \bar{\hat{G}}\!\left( x \right) \right]}^3}
\right\}   
 $}, 
\end{align}
where the dotted accent for $\dot{Q}\left(\left| \tilde{\alpha } \right|\right)$ is defined by the modified derivative i.e., $\dot{Q}\!\equiv\!(\!1/|2\tilde{\alpha }|)dQ/{d\left| \tilde{\alpha } \right|}$, and $\hat{G}{{\left( x \right)}_{ab}}$ is defined as:
\begin{align}\label{eqB10}
\scalebox{0.76}{$\displaystyle
\hat{G}{\left( x_1,x_2 \right)}_{ab}\!\equiv\!\left\langle{{\hat{\psi }}_a}\left(x_1\right){{\hat{\psi }}_b}\left( x_2 \right) \right\rangle\!\!\equiv\!\!\left(1\!+\!\frac{1}{2\tilde{d}}g^{\mu \nu }{\bar{\nabla }}_{\mu }{\nabla }_{\nu } \!\right)\!{G_{\psi }}{{\left( x_1,x_2 \right)}_{ab}}
 $}\nonumber\\
\scalebox{0.76}{$\displaystyle
\bar{\hat{G}}\left( x \right)\equiv{\psi }_a\hat{G}{{\left( x \right)}_{ab}}{\psi }_b,\;\mathrm{Tr}\hat{G}\left( \mathbf{0} \right)\equiv \sum\limits_{l}{\hat{G}{{\left( \mathbf{0} \right)}_{ll}}}
 $},\quad\quad  
\end{align}
where the trace refers only to the index of the $2\times 2$ matrix $\hat{G}{{\left( x \right)}_{ab}}$. To calculate the free energy of the two-loop correction, introduce the hopping expansion of $\hat{G}{{\left( x \right)}_{ab}}$, as follows:
\begin{align}\label{eqB11}
\scalebox{0.82}{$\displaystyle
\hat{G}{{\left( x,y \right)}_{ab}}=\frac{{\tilde{b}}}{{\tilde{d}}}\sum\limits_{n=0}^{\infty }{\left( \frac{\tilde{b}\eta }{{\tilde{d}}} \right)}_{ab}^{n}h{{\left( x,y \right)}^{n+1}},
 $}\nonumber\\
\scalebox{0.82}{$\displaystyle
h\left( x,y \right)\equiv 2\tilde{d}{{\delta }_{xy}}+{g^{\mu \nu }}{{\bar{\nabla }}_{\mu }}{{\nabla }_{\nu }},
 $},\quad\quad  
\end{align}
For $\tilde{b}\ll\tilde{d}$, i.e., at the limit of small ${{E'}\!_S}$, the free energy due to the mean field approximation with up to the two loop corrections for the up to order ${{\tilde{b}}^4}$ is as follows:
\begin{align}\label{eqB12}
\scalebox{0.7}{$\displaystyle
{F'}=\frac{{{\tilde{\psi }}^2}}{4\tilde{b}}
-\log {I_0}\left( \alpha  \right)
-\frac{{{\tilde{b}}^2}}{2\tilde{d}}\left( \eta _L^2+\eta _{T}^{2} \right)\left( 1+\frac{1}{2\tilde{d}}{{{\hat{h}}}_2} \right)
$}\quad\quad\quad\quad\quad\nonumber\\
\scalebox{0.7}{$\displaystyle
\quad\quad-\frac{{{\tilde{b}}^3}}{6{{{\tilde{d}}}^2}}\left[ 3{{\left( {{{\dot{\eta }}}_T} \right)}^2}+{{\left( {{{\dot{\eta }}}_L} \right)}^2} \right]
-\frac{{{\tilde{b}}^4}}{2{{\tilde{d}}^2}}\left\{ \left( 3\eta _{T}^4+3\eta _L^4 \right)\left( 1-\frac{1}{2\tilde{d}}+\frac{{{\hat{h}}_4}}{12{{\tilde{d}}^2}} \right)
\right.
 $}\nonumber\\
\scalebox{0.7}{$\displaystyle
\left.
+\frac{1}{{\tilde{\psi }}}\left( 3{{{\dot{\eta }}}_T}\eta _T^2-4{{\eta }_L}{{\eta }_T}{{{\dot{\eta }}}_T}+2{{\eta }_L}{{\dot{\eta }}_L}{{\eta }_T} \right)+\eta _L^2{{{\ddot{\eta }}}_L} \right\}+0\left( {{{\tilde{b}}}^5} \right)
 $},\quad\quad\nonumber\\
\scalebox{0.7}{$\displaystyle
 {{\hat{h}}_{2}}\equiv 2\left( {{{\tilde{\gamma }}}^2}-\tilde{\gamma } \right),\; {{\hat{h}}_4}\equiv 6{{\tilde{\gamma }}^4}+\left( 24D-12 \right){{\tilde{\gamma }}^{2}}+\left( 6-24D \right)\tilde{\gamma }
 $},\quad 
\end{align}
The result of finding the minimum value ${{\tilde{\psi }}_{0}}$ of $\tilde{\psi }$ according to Eq.(\ref{eqB12}) is as follows:
\begin{align}\label{eqB13}
\scalebox{0.78}{$\displaystyle
{{\tilde{\psi }}_0}=\sqrt{\frac{8\left( 1-{1}/{{\tilde{b}}}\; \right)+32{{\Delta }_2}}{1-64{{\Delta }_1}}}
$},\quad\quad\quad\quad\quad\quad\nonumber\\
\scalebox{0.78}{$\displaystyle
{{\Delta }_{1}}\equiv \frac{{{{\tilde{b}}}^4}\left( 580{{{\tilde{d}}}^2}-612\tilde{d}+51{{{\hat{h}}}_4} \right)+68{{{\tilde{b}}}^2{{{\tilde{d}}}^2}}\left( 2\tilde{d}+{{{\hat{h}}}_2} \right)}{4096{{{\tilde{d}}}^4}}
 $},\quad\nonumber\\
\scalebox{0.78}{$\displaystyle
{{\Delta }_2}\equiv \frac{38{{{\tilde{b}}}^4}{{{\tilde{d}}}^2}-60{{{\tilde{b}}}^3}{{{\tilde{d}}}^2}\left( 2\tilde{d}+{{{\hat{h}}}_2} \right)-3{{{\tilde{b}}}^5}\left( 34{{{\tilde{d}}}^2}-60\tilde{d}+5{{\hat{h}}_4} \right)}{768{{\tilde{d}}^4}\tilde{b}}
 $},\; 
\end{align}
Therefore, for the mean field approximation with up to the two loop corrections of ${E'}\!_S$, the critical point $\left( {E'}\!_S \right)_c^{2\text{loop}}$ and the triple critical point ${{\left( {E'}\!_S \right)}^{t\!r\!i}}$ are respectively as follows:
\begin{align}\label{eqB14}
\scalebox{0.62}{$\displaystyle
\left( {{{E'}}_S} \right)_c^{2\text{loop}}=\frac{1}{\left( D+\gamma  \right)}\Bigl\{ 1-\Bigl[120{D^3}+8{D^2}\left( 15{{{\tilde{\gamma }}}^2}+30\tilde{\gamma }+8 \right)
\Bigr.\Bigr.
 $}\quad\quad\quad\quad\quad\nonumber\\
\scalebox{0.62}{$\displaystyle
\Bigl.\Bigl. 
+4D\left( 60{{{\tilde{\gamma }}}^3}+120{{{\tilde{\gamma }}}^2}-58\tilde{\gamma }-45 \right)+2\gamma \left( 105{{{\tilde{\gamma }}}^{3}}-58\tilde{\gamma }-45 \right)\Bigr]/{192{{\left( D+\tilde{\gamma } \right)}^4}}
 \Bigr\}^{-1}
 $},
\end{align}
\begin{align}\label{eqB15}
\scalebox{0.65}{$\displaystyle
{\left( {E'}\!_S \right)}^{t\!r\!i}\!=\!\sqrt{\frac{\sqrt{1156{{\left( 2\tilde{d}+{{\hat{h}}_2} \right)}^2}\!+\!64\left( 580{{\tilde{d}}^2}\!-\!612\tilde{d}\!+\!51{{\hat{h}}_4} \right)}\!-\!34\left( 2\tilde{d}\!+\!{\hat{h}}_2 \right)}{580{{\tilde{d}}^2}\!-\!612\tilde{d}\!+\!51{{\hat{h}}_4}}} 
 $}, 
\end{align}

\end{document}